%% file: main.tex
\documentclass[journal, onecolumn]{ieeetran}
\usepackage[utf8]{inputenc}
\usepackage{amsmath,amssymb,mathrsfs,amsthm, amsfonts}
\usepackage{bbm}
\usepackage{chemarrow}
\usepackage{graphicx}
\usepackage{xcolor}
\usepackage{accents}
\usepackage{sidecap}
\usepackage[colorlinks=true,allcolors=blue]{hyperref}
\usepackage{breakurl} 
\usepackage{soul}
\usepackage{float}
\usepackage[shortlabels]{enumitem}
\usepackage{epstopdf}
\usepackage{footnote}
\usepackage{algorithmic}
\usepackage{algorithm}
\usepackage{array}
\usepackage[caption=false,font=normalsize,labelfont=sf,textfont=sf]{subfig}
\usepackage{textcomp}
\usepackage{stfloats}
\usepackage{url}
\usepackage{verbatim}
\usepackage{cite}
\usepackage{mathtools}

\input{macros}

\begin{document}

% title
\title{
The Geometry of Statistical Data and Information: A Large Deviation Perspective
}

\author{
\IEEEauthorblockN{Viswa Virinchi Muppirala\IEEEauthorrefmark{1} and Hong Qian\IEEEauthorrefmark{2}}\\[6pt]
\IEEEauthorrefmark{1}Department of Electrical and Computer Engineering, University of Washington, Seattle, WA 98195, USA,\\
email: virinchi@uw.edu\\[6pt]
\IEEEauthorrefmark{2}Department of Applied Mathematics, University of Washington, Seattle, WA 98195, USA,\\
email: hqian@uw.edu
}

\vspace{1cm}

%%%%%%%%%%%%%%%%%% TITLE %%%%%%%%%%%%%%%%%%%%%

\maketitle

%%%%%%%%%%%%%%%%%% ABSTRACT %%%%%%%%%%%%%%%%%%

\begin{abstract} 
The manifold of empirical mean values of statistical data {\em ad infinitum} has a geometric shape that depends on the probability measure that governs the generating model. Large deviation theory produces entropy functions that depend on both the probability measure and the statistical data; we use entropy to study the geometry of the data space rather than that of the space of probability distributions. It is well known, since Rao's work, that the Fisher-Rao metric makes the probability simplex into a sphere. From our perspective, that result translates to the space of empirical singleton counting frequencies under an i.i.d. assumption. Following our ideas and going beyond i.i.d., the choice of measure curves the space. When we study the pairwise statistics, the spherical geometry breaks down entirely. We show that the information projection, defined in information geometry as divergence minimization, coincides with the information projection in Kolmogorov's probability theory.  This identification holds under both i.i.d. and Markovian assumptions and connects information geometry to the foundations of probability theory.

\end{abstract}

%%%%%%%%%%%%%%%%%% INTRO %%%%%%%%%%%%%%%%%%

\section{Introduction}

There is a deep mathematical kinship between the modern theory of probability \cite{KolmogorovBook} and differential geometry as practiced in theoretical physics \cite{choquet-bruhat-at-al}. An elementary random event represents a complex real-world phenomenon---whether an object, a process, or both, and according to the former, their numerical expressions are called random variables. According to the latter, events and processes are identified with a geometric entity whose numerical representations are coordinate (chart) dependent; the mathematical {\bf\em model} of a phenomenon or phenomena itself needs to be a vector or a tensor that is coordinate independent (gauge invariant). These geometric concepts are inherent in {\em linear algebra}, which studies abstract mathematical objects in $\mathbb{R}^n$: $n$-tuples and $n\times n$ matrices of numbers are merely coordinates of vectors and linear transformations; they are dependent upon chosen bases and are not intrinsic.  

Information geometry (IG) ~\cite{amari-book,Amari1990-ta} studies the geometric structure inherent to the space of probability distributions as family of statistical models: It is a mathematical theory of inter-relations among different statistical models. It interprets ``information'' in empirical data through the lens of probability that ultimately underlie all statistics. The primary object of IG is the Fisher information metric generated by divergence functions; they are used to quantify the difference between two probabilistic models as geometric {\em distance}. C. R.  Rao was one of the first to point out that the Fisher information matrix originated in statistical inference can be used as a Riemannian metric to measure a symmetric difference between two probability distributions ({\em i.e.}, two models). He~\cite{Rao1992-qj} recognized similarities between the properties of the Fisher information matrix and that of a Riemannian metric: its symmetry, positive definiteness, and behavior as a second-order covariant tensor. 

Embracing the Fisher information matrix as a Riemannian metric, S.-I. Amari developed the {\em geometry of the space of probability distributions}, which, in more advanced mathematics, could be thought of as the space of probability measures \cite{zhangjun}. 
He focused on manifolds that possess a global chart.   When such a global chart is absent, he confined his discussions to a local patch of the statistical manifold~\cite{Amari1990-ta}. With the global chart available from statistical applications, the nature of an applied geometric investigation differs from that of more abstract mathematics: establishing a Riemannian metric amounts to identifying a {\em scalar potential function} whose local Hessian matrix represents the geometric metric.  The Hessian provides a special bases for the Fisher metric with the statistical log-likelihood function as a potential.  The manifolds with the Fisher metric are generally not flat under the natural Levi-Civita connection.  
{{Here, it is important to point out that the notion of convexity of a function depends on the choice of a connection while bypassing the choice of the metric.}}
Amari discovered a hidden ``flat'' geometry within statistical manifolds by introducing non-Levi-Civita connections via a convex dual that is flat. This insight into the differential geometry of statistical manifolds revealed that they could be dually flat when examined through convex duality and their dual connections. It leads to the establishment of a generalized Pythagorean theorem based on a Bregman divergence. The book \cite{amari-book} presents a wide range of applications of this approach across various fields, including machine learning, statistical inference, signal processing, economics, and neuroscience.

{ We study the geometry of the space of statistical data rather than the geometry of the space of probability distributions. Large deviation theory, with the rate function as the primary object, provides the natural framework for this investigation and justifies the choice of metric through duality with the cumulant generating function.}
 In LDT, the space of probability measures equipped with the Wasserstein distance is a Polish space (separable, completely metrizable topological space) \cite{Ambrosio2008, Villani2009,zhangjun}; it offers a rigorous framework for discussing the space of all probability measures, even for the space of continuous measures where the dimension is infinite.
 The notion of information projection appears independently in two bodies of literature, with different motivations and distinct mathematical objects. Information projection, in the literature on information geometry \cite{csiszar1975divergence, csiszar2003information}, operates in the space of \emph{probability distributions} and defines the projection by minimizing the KL divergence. Probability theory's information projection \cite{durrett2019probability,van2007stochastic} operates on random variables and $\sigma$-algebras and defines projection via an $L^2$ minimization.  The present work unifies these two notions by studying the geometry of statistical data as \emph{random variables}, a perspective that was missing in the information geometry literature. 
 % We capture the fundamental depth of information projection as described in probability theory emerging from $\sigma$-algebras and measurements.
 This refinement improves Amari's connections of the conditional expectation step to information projection (also known as exponential projection) in his application of information geometry in EM algorithm \cite{AMARI1995EM}.

In developing a geometric framework for information, it is crucial not to confine ourselves solely to the study of probability distributions or statistical manifolds; entropy is an integral part of IG. E.T Jaynes ~\cite{JaynesBook} emphasizes that probabilities should be viewed as an extension of logic when deductive reasoning falls short due to incomplete information. While Probability and Bayesian logic do provide frameworks for inferences, we must also recognize empirical frequencies as meaningful scientific/engineering measurements on recurrent dynamical systems in the real world, which researchers routinely represent in non-linear physics and applied mathematics \cite{strogatz-book, wellner-book-2, QianGeBook}. 

Normalized empirical measures, empirical frequency for short, and probabilities occupy the same mathematical space, yet they differ fundamentally in their scientific nature. Empirical frequencies represent observed outcomes—statistics from actual data collected from measurements, even imagined as a thought experiment ({\em Gedankenexperiment}). In contrast, one best understands probabilities by adopting the Bayesian perspective as degrees of belief informed by theoretical modeling and prior knowledge. While it is entirely sensible to inquire about the occurrence of specific empirical frequencies, posing similar questions about probabilities does not make sense, as probabilities are theoretical and not directly observable as in principle. Only in the mathematical limit are theoretical objects such as probabilities and expected values related to empirical measurements with confidence---the philosophical significance of the Law of Large Numbers. 

This paper identifies entropy as the rate function in large deviation theory, which is entirely consistent with the entropy in thermal physics \cite{callen-book,qian_entropy_24}. These rate functions arise in the limit of data \textit{ad infinitum} and will serve as a foundation for our geometrical exploration of information \cite{mqw_paper_1}. It is fitting here to quote P. W. Anderson (1923-2020), a leading theoretical physicist who defined the statistical physics of emergent behavior \cite{anderson72}:
\begin{quote}
{\em ``Starting with the fundamental laws and a computer, we would have to do two impossible things | solve a problem with infinitely many bodies, and then apply the result to a finite system | before we synthesized this behavior.''
}
\end{quote}
Analogous to statistical thermodynamics, we begin with a simple state space and count the number of large sequences. Shannon's entropy is a rate function for increasing multiplicity. As we incorporate further assumptions like probability with identical and independence, Kullback-Leibler divergence (KLD) also emerges as a rate function. Sanov's theorem, in particular, establishes the existence of an almost universal convex function, which we will identify as the Entropy function and as a divergence for information theory and in IG, respectively. This rate function arises in the limit of data \textit{ad infinitum} and will serve as a foundation for our geometrical exploration \cite{mqw_paper_1}. As a mathematical limit, it is non-random. This entropy function not only serves as the basis for the divergence functions central to IG but also closely aligns with the theory of statistical thermodynamics, where entropy is traditionally defined as the logarithm of the number of arrangements of a system. Legendre-Fenchel duality plays a prominent role in this regard. Given a prior probability, our approach treats the Entropy function as a function of empirical mean values from repeated measurements. With the expectation implied by the prior probability, it is a bivariate function of ``data'' and ``model''.   This treatment is our interpretation of the divergence function in IG. The entropy function in Thermodynamics is a Legendre-Fenchel dual of a free energy function. In large deviation theory, rate functions can have closed forms as the Legendre-Fenchel transform (LFT) of the cumulant-generating function,  as seen in Cram\'{e}r's theorem and its generalization G\"{a}rtner-Ellis theorem. LFT makes contraction particularly simple \cite{qian-cheng-qb} with linear additivity \cite{angelini_qian}.

This paper studies the geometry of statistical data and how the underlying model defines that geometry. We consider recurrent data and study empirical statistics based on a large number of observations. In this context, a large deviation theory rate function, a function of both data and model, exists and is our central object. We justifiably use the covariance of statistical measurements to define a metric in the space of statistical data. 
Under the i.i.d.\ assumption, the space of empirical counting frequencies possesses a spherical geometry that is independent of the underlying probability measure $\vp$. When the data are Markovian-dependent, the geometry of empirical frequencies explicitly depends on the transition matrix $\mP$. Knowledge of singleton counts alone no longer determines pairwise dependence. In this setting, probability theory functions extends logic, filling in missing structural information through the model. In the Markovian case, the model instructs the data on how to curve. The geometry of statistical data is therefore not intrinsic to the counts alone; it is shaped by the structural assumptions imposed on the generative process. 

The projection of a point $q$ in a space $\mathcal{M}$ onto a subspace $\mathcal{U}$, as taught in elementary courses, is often seen as the point in $\mathcal{U}$ that is at the minimal distance from $q$. Information projection, on the other hand \cite{csiszar1975divergence, nielsen2018information, amari-book} uses a dissimilarity measure rather than a distance. Probability theory provides a clear definition of the $L^p$ distance between measurable random variables. In this work, we utilize the $L^2$ distance to conclude that information projection through probability theory under i.i.d and Markovian measure coincides with the definition of information projection and Markov information projection \cite{csiszar1987conditional} in the literature on information geometry.

\subsection{Organization of this work}
We organize our main results of the paper as follows. 
\begin{itemize}
    \item In Sec. \ref{sec:2} we define empirical frequencies, probabilities and, emphasize their distinction. We introduce necessary concepts from large deviation theory, and describe the information-theoretic nature of rate functions. In Sec. \ref{sec:probtheorey-infoproj}, we define information gain and information projection in the language of probability theory.

    \item In Sec. \ref{sec:3} we show how LDT generates rate functions as information-theoretic quantities. In connection with Gibbs' method of thermodynamics and LFT, we provide an energetic description of i.i.d. data. In Sec. \ref{subsection:contraction}, we provide a principled justification for the metric using the ideas from nondimensionalization. These pave the way for one of our key insights in Sec. \ref{subsection:info-proj}, we show that information projection, central to information geometry, agrees with information projection from probability theory.  We connect information geometry's Pythagorean theorem through conditional expectation and information gain.  This result unifies the notions of ``projection'' in the two different settings rigorously. 

    \item More importantly, carrying our approach further beyond the i.i.d. case, we study identically distributed but Markov-dependent data in Sec. \ref{sec:3_markov}, \ref{cgf}, \ref{sec:empfreqgeom}, and \ref{sec:curvature}. 
    
    \item Sec. \ref{sec:4} contains our conclusion. 
\end{itemize}

%%%%%%%%%%%%%%%%%% BACKGROUND AND Preliminaries %%%%%%%%%%%%%%%%%%
\section{Preliminaries and background}
\label{sec:2}

Fix a finite  set $\mathcal{S} = \{1, 2, \ldots, n\}$ called a state space. We introduce our notation to describe random variables on $\mathcal{S}$.

\begin{notation}{(Random variable)} \label{rv-matrix}
    For a positive integer $k \in \mathbb{Z}_{>0}$, consider $k$ real-valued ($\reals$-valued)  random variables or equivalently, a $\reals^k$-valued random variable $\vX \colon \mathcal{S} \to \reals^k$ taking values $\vx_1, \ldots, \vx_n \in \reals^k$. We represent the random variable $\vX$ in matrix-form with the values $\vx_1, \ldots, \vx_n$ as columns 
    \[\vX  = \begin{bmatrix}
    | & | & & | \\
    \vx_1 & \vx_2 & \cdots & \vx_n \\
    | & | & & |
    \end{bmatrix}.\]
    
    Here $\vX$ is a $k \times n$ matrix. Note that each row of the matrix $\vX$ is a real-valued random variable.
\end{notation}

We explore this notation in the following example.
\begin{example}{(Indicator random variables)} \label{ex:indicator}
    Consider the $n$-dimensional identity matrix 
    \[
    \mI_n = \bmat{
    1 & 0 & \cdots & 0 \\
    0 & 1 & \cdots & 0 \\
    \vdots & \vdots & \ddots & \vdots \\
    0 & 0 & \cdots & 1
    }.
    \]
    The rows of $\mI_n $ denotes $n$ indicator random variables $\{\indicator_1,\ldots,\indicator_n\}$. For $i \in \mathcal{S}$, we define each indicator random variable  $\indicator_i \colon \mathcal{S} \to \{0,1\}$ by
    \begin{align*}
        \indicator_i(j) = \delta_i^j,
    \end{align*}
    where $\delta_i^j$ is a Kronecker delta function that returns $1$ if $i$ and $j$ are equal and $0$ otherwise. 
\end{example}

We define empirical counting frequencies or simply empirical frequencies as follows.

\begin{definition}{(Empirical frequencies and counts)}\label{empirical-freq}
    Fix a positive integer $N \in \mathbb{Z}_{>0}$ and consider $N$ samples from $\mathcal{S}$. That is, $\omega \in \mathcal{S}^N$ is a $N$-length sequence and the $ j$-th symbol in $\omega$ is $\omega(j)$. We define empirical counts as occurence counts of elements in $\mathcal{S}$. We define  empirical frequencies as  normalized occurrence counts of elements in $\mathcal{S}$. For any $i \in \mathcal{S}$, the empirical frequency of $i$ is the count of occurrences of $i$ in the $N$ samples divided by $N$.

    \begin{align*}
        c_i &= \text{ \# occurrences of $i$ in $N$ samples} = \sum_{j=1}^N \indicator_i(\omega(j))  \\
        \nu_{i} &= \frac{c_i}{N}.
    \end{align*}

    We use the vector $\vnu = \bmat{\nu_1 &\nu_2 & \ldots & \nu_n}^T $ to denote the empirical frequencies and the vector $\vc = \bmat{c_1 &c_2 & \ldots & c_n}^T  $ to denote empirical counts. Note that both empirical frequencies and counts are random variables by definition. And moreover, in vector notation
    \[
\vc = \sum_{j=1}^N \mI_n^{(j)}, \vnu = \frac{1}{N} \sum_{j=1}^N \mI_n^{(j)}.
\]

\end{definition}
\begin{notation}
     We introduce the notation $\ones_n$ to denote the $n$-vector with all ones, i.e., $\ones_n = \bmat{1&1 & \ldots & 1}^T $ where $\ones_n \in \mathbb{R}^n$.
\end{notation}

\begin{definition}{(The space of empirical frequencies)}\label{simplex}
    One may be familiar with the definition of probability simplex in $\reals^n$ as
    \[
    \Delta^{n} = \left\{ \boldsymbol{\nu} \in \mathbb{R}_{\geq 0}^n \bigg| \sum_{i=1}^n \nu_i = 1 \right\}.
    \]
    
    Similarly, we define the space of empirical frequencies and denote it as $\riDelta{n}$. It is the relative interior of $\riDelta{n}$:
    \[ \riDelta{n} = \Big\{ \vnu \in \reals^n_{> 0} \Big| \sum_{i=1}^n \nu_i =1 \Big\} =  \Big\{ \vnu \in \reals^n_{> 0} \Big| \ones_n^T \vnu =1 \Big\}\]
    
    Even though the individual counting frequency can be zero, we only focus on positive counting frequencies. 

    For a positive integer $N$, the $N$-\emph{lattice} of the simplex is
    \[
\Delta^{n}_{N} \;=\; \Big\{ \boldsymbol{\nu}\in \Delta^{n} \;:\; N\boldsymbol{\nu}\in \mathbb{Z}_{>0}^{\,n} \Big\}.
\]
\end{definition}

\subsection{History of entropy and revisiting Shannon's work}

Ludwig Boltzmann \cite{boltzmann1872-english}, in his kinetic theory of gasses, was the first person to relate entropy ($H$) to the number of possible configurations ($W$). Max Planck \cite{planck1914theory} formally reduced it to the equation we now know as
\begin{align}\label{maxplanck-formula}
    S = k_B \log W.
\end{align}

Planck defines $k_B$ as a universal constant, which is now known as the Boltzmann constant. Planck and Boltzmann consider an example of $N$ molecules in a volume divided into $n$ spacial elements. They explore the combinatorial problem of finding the number of arrangements of $N$ molecules in $n$ spacial elements with counting frequencies $\vnu$. This problem is equivalent to finding the number of sequences of length $N$ with counting frequencies $\vnu$. The number of sequences follows the multinomial coefficient 

\begin{align}
    \label{multinomial}
     W(N,\vnu)= {N\choose N\nu_1\ N\nu_2\ \cdots\  N\nu_n}.
\end{align}

They consider the case of large $N$ and use Stirling's formula $n! = \sqrt{2 \pi n}\left(\frac{n}{e}\right)^n$. Max Planck plugs it into equation (\ref{maxplanck-formula}) and finds the entropy as the function of counting frequencies \footnote{Upon examining Planck's work, one might notice that he frequently uses the term `probability' to refer to what we describe as counting frequencies. He often refers to $W$---the number of arrangements as a measure of probability, suggesting configurations with higher counts are more probable. This interpretation arises because Boltzmann and Planck's work predates the formal treatment of probability theory by Kolmogorov in 1938 \cite{KolmogorovBook}.} as
\begin{align}
    S = -k_B N \sum_{i=1}^n \nu_i \log \nu_i.
\end{align}

The relation between entropy and the logarithm of the number of configurations is fundamental and intuitive. In information theory, this perspective is particularly practical$\colon$ knowing that the number of configurations is $W$, we require $\lceil \log_2 W \rceil$ bits to represent each configuration. Alternatively, one can view this as a variation of the game ``20 questions''; suppose a person chooses one of the $W$ sequences, and another person has to guess the correct sequence after asking a series of $k$ yes or no questions. We can ask about the minimum value of $k$ to know the sequence with absolute certainty, and the answer is $\lceil \log_2 W \rceil$. 

The present work looks at entropy as the asymptotic limit of infinite sampling, where the concept appears as a growth rate ``per data'',  again as a ``derivative'' of $\infty$ divided by $\infty$. The number of possible sequences $W(N, \vnu)$ in equation (\ref{multinomial}) grows exponentially as $N$ increases and we calculate the rate of growth for large $N$ using the limit

\begin{align}\label{eq:shannon-entropy-freq}
     \lim_{N \to \infty} \frac{\log W(N, \vnu)}{N} = \lim_{N\to\infty} 
    \frac{1}{N}\log {N\choose N\nu_1\ N\nu_2\ \cdots\  N\nu_n} = -\sum_{i=1}^n \nu_i \log \nu_i.
\end{align}

As a mathematical concept, this formula for entropy emerges from the limit of data {\em ad infinitum}. It also allows us to interpret it as the rate at which information grows as we gather more observations.

C. Shannon, in his paper ``A Mathematical Theory of Communication''\cite{shannon1948mathematical}, was interested in ergodic sources producing symbols and focused on the rate of information produced as a function of stationary probabilities $\vp = \bmat{p_1 & p_2 &\ldots p_n}^T$. Shannon proposed that this function should follow certain axioms and found that $H(\vp)= -\sum_{i=1}^n p_i\log p_i$ up to a multiplicative constant uniquely satisfies these properties. So, he focuses on the entropy of a source rather than the entropy derived from observed sequences, and hence, his formula is a function of probabilities $\vp$.

Subsequently, Kinchin \cite{khinchin2013mathematical} demonstrated a similar result based on his own axioms, and Shore and Johnson \cite{shore-johnson}, inspired by E.T. Jaynes's Maximum Entropy Principle, proved that cross-entropy or relative entropy is the unique function fulfilling their own specific axioms. All the discussed works employ probability theory and an axiomatic approach as their foundation.

In connection to probability, if one supposes the symbols in a sequence are from a stationary with identical, not necessarily independent, probability distribution $\vp = \bmat{p_1 & p_2 &\ldots p_n}^T$ for the corresponding states $\mathcal{S}$, then in the asymptotic limit of $N\to\infty$, the ergodic theorem states that the observed empirical frequencies $\vnu\to\vp$, and thus one has the asymptotic rate of counting in \eqref{eq:shannon-entropy-freq} becoming
\begin{equation}
\label{eq:shannon-2}
 -\sum_{i=1}^K p_i\log p_i.
\end{equation}
The previous expression represents the asymptotic rate of the number of {\em typical sequences}, which we roughly define as the sequences whose empirical frequencies $\vnu$ equal $\vp$.

In his paper \cite{shannon1948mathematical}, Shannon highlights that his axiomatic definition is ultimately the ``rate
of growth of the logarithm of the number of reasonably probable sequences''.  
He considers an ergodic source that generates a sequence of length $N$ with probabilities $\vp$. For a probability $0<\epsilon<1$, he defines $M(N,\epsilon)$ as the number of sequences required until the cumulative probability of observing the sequences reaches $\epsilon$. He recognizes his entropy $-\sum_{i=1}^n p_i \log p_i$ as the rate function of $M(N,\epsilon)$. That is,

\begin{align*}
    \lim_{N \to \infty} \frac{\log M(N,\epsilon)}{N} = -\sum_{i=1}^n p_i \log p_i.
\end{align*}

 Shannon initially terms his entropy as a characteristic of the source and then relates it to the statistic of the sequences it generates. We argue that the empirical, frequency-based form of entropy is broader and more versatile — it directly relates entropy to the logarithm of the number of possible states via simple counting arguments.

 It does not assume any specific source for the sequence, ergodicity, or even probabilities. This description also aligns closely with its usage in thermodynamics as a fundamentally statistical concept rather than one rooted in probability. 

If we further assume that the symbols in a sequence are independent and identically distributed (i.i.d.), we can also interpret Shannon's entropy as a rate of vanishing randomness as $N\to\infty$. Thus, it is related to the growth of one's confidence and {\em certainty}. 

The probability of observing a particular sequence with the empirical frequencies $\vnu \in \Delta^n_N$ is:
\begin{align}
\label{eq:vanish}
\prod_{i=1}^n    p_i^{N\nu_i} = \exp\Big\{\sum_{i=1}^n N \nu_i \log p_i\Big\}.
\end{align}
This expression for the probability of $\vnu$ is also a function of the probabilities $\vp=(p_1,\cdots,p_n)$.  One may understand the latter as a set of parameters for the most general statistical model for $\mathcal{S}$. For a given sequence observed with $\vnu$ fixed, one could ask: for which set of $\vp$ is the rate of vanishing randomness minimized, {\em e.g.}, the probability maximized? This is a maximum likelihood estimation (MLE) problem. We seek the probability $\vp$ as a set of parameters that maximizes the likelihood of observing the particular sequence. The likelihood function is the expression in \eqref{eq:vanish}. Taking the logarithm of the likelihood function, we obtain the log-likelihood:
\[
\log L(\vp) = \sum_{i=1}^n N\nu_i \log p_i,
\]
we differentiate $\log L(p_1, \dots, p_n)$ w.r.t each $p_i$ and set the derivative equal to zero, subject to the constraint $\sum_{i=1}^n p_i = 1$. The solution yields $p_i = \nu_i$, for which the rate of vanishing randomness is  $-\sum \nu_i \log \nu_i$, which is again, Shannon's entropy. One may also use Jensen's inequality to confirm $\sum_{i=1}^n\nu_i\ln p_i\le \sum_{i=1}^n\nu_i\ln\nu_i$. 

Combining our previous discussions on counting the number of sequences with a fixed $\vnu$ and the rate of vanishing randomness for a sequence with composition $\vnu$, one might ask the question: what is the probability of observing a sequence with empirical frequencies $\vnu = \vq  \in \Delta_N^n$? This leads to multiplying the counting term and the probability term, which gives us:

\begin{align*}
     \Prob[\vnu  = \vq]  = e^{-N \sum q_i \log q_i} \times e^{N \sum q_i \log p_i} = e^{-N \sum q_i \log \left(\frac{q_i}{p_i}\right)},
\end{align*}

where we recognize $\sum q_i \log \left(\frac{q_i}{p_i}\right)$ as the KLD, a function of both the observed empirical frequencies  $\vq$ and the probability distribution defined by $\vp$.

In a general setting, where $\vq \in \Delta^n$ we instead consider an open ball around $\vq$ and denote it as $\vnu \in \rd \vq$. For such sets, the asymptotic rate for $\Prob [\vnu \in \rd \vq]$ yields \cite{dembo-book}
\begin{align}
    \label{sanov-rate}
        \log \Prob\!\big[\vnu \in \rd \vq\big]
    \approx   -N\,H(\vq \mid \vp).
\end{align}

Eq. (\ref{sanov-rate}) is a variant of Sanov's theorem, and we recognize the KLD as the rate function of Sanov's theorem \cite{Sanov} \cite{dembo-book}: The rate of vanishing probability for all $\vnu \neq\vp$, which, of course, implies the growth of certainty for $\vnu=\vp$.  

Interestingly, we can also explain Eq. (\ref{sanov-rate}) in connection to the Principle of Maximum Log-Likelihood: 
\[
   \sum_{i=1}^n q_i\log p_i
   \le \sum_{i=1}^n q_i\log q_i,
\]
which identifies $\vq$ as the ``best'' $\vp$ for the observed $\vq$.  This suggests that the function in (\ref{sanov-rate}) has dual roles, in the theory of probability as a negative large deviation rate function and in the theory of statistics as a log-likelihood function, treating $\vp$ as a set of parameters.

We, therefore, found important meanings for all three
\[
  -\sum_{i=1}^n q_i\log q_i, \ 
  -\sum_{i=1}^n p_i\log p_i,  \text{ and } \, 
   \sum_{i=1}^n q_i\log p_i.
\]

This document often refers to KLD or any rate function involving empirical measurements as entropy. We now define entropy for empirical counting frequency.

\begin{definition}{(Entropy for empirical counting frequency)}
    We define entropy $H(\vq \mid \vp)$ as the rate function of asymptotic probability in Sanov's large deviation theorem represented in equation (\ref{sanov-rate}). 

    \begin{align*}
        H(\vq \mid \vp) =  \sum q_i \log \left(\frac{q_i}{p_i}\right) = D_{\text{KL}}\left( \vq \| \vp \right).
    \end{align*}
\end{definition}

The previous discussion lays the foundation for empirical frequencies; we observe their significance as the first argument in the KLD. This independent variable represents frequencies derived directly from data observations, which differ sharply from the second argument, comprising probabilities as our model's parameters. These probabilities are mathematical constructs designed to model our beliefs and prior knowledge.

\subsection{Information gain and information provided by a random variable: Information projection}
\label{sec:probtheorey-infoproj}

{{ 
So far, we have not involved the proper space of events: a $\sigma$-algebra. Unrelated to the earlier defined state space $\mathcal{S}$, to understand the information provided by repeated measurements of a random variable, one needs to consider a finite measurable space $\left(\Omega, \mathcal{F}\right)$.  Even for the simplest i.i.d. sequence for i.i.d. samples the $\Omega=\mathcal{S}^N$ and $\mathcal{F}=2^{\Omega}$.

\paragraph{Information gain} Even in the absence of a probability measure, a measurable space $(\Omega, \mathcal{F})$ exhibits a rich structure that provides a natural analytic framework for information.  Intuitively, information is acquired by drawing logical conclusions, and a $\sigma$-algebra embodies an inherent Boolean algebraic structure.}} Whether the space is described in terms of points or in an abstract fashion \cite{kappos2014probability, rota2001twelve}, the $\sigma$-algebra $\mathcal{F}$ is endowed with a natural partial order: if $A,B \in \mathcal{F}$ with $A \subset B$, then the occurrence of event $A$ is inherently more informative than that of event $B$. We may define the quantity of additional information provided by $A$ relative to $B$ as $\log\left(\frac{|B|}{|A|}\right)$, where $|\cdot|$ denotes the size of the event as a subset.  The probability measure $\mathbb{P}$ is the ``size'' defined on $\mathcal{F}$, the information gain then is given by \cite{kolmogorov-information}
\[
\log\left(\frac{\mathbb{P}[B]}{\mathbb{P}[A]}\right).
\]
In terms of $\sigma$-algebras, if $\mathcal{G}$ is a refinement of $\mathcal{F}$ (i.e., $\mathcal{G} $ is finer than $ \mathcal{F}$), then $\mathcal{F}$ contains no more information than $\mathcal{G}$.

\begin{example}
\label{example:infogain1}
    We continue to consider $\Omega=\mathcal{S}^N$ of all $N$-length sequences in $\mathcal{S}$. Let $A \subset \Omega $ be the space of all sequences with empirical counting frequency $\vnu \in \Delta_N^n
$. Then, for a large $N$, 
without involving a probability measure, we obtain approximately  $\log \frac{n^N}{W(N,\vnu)} = N S \left(\vnu \, \mid \, (1/n, \ldots, 1/n) \right)$ more information knowing the empirical frequency.

\end{example}

\paragraph{Information projection}  Consider a random variable $X \colon \Omega \to \reals$. This random variable generates a unique sub-$\sigma$-algebra we usually call $\sigma(X)$, but we will use the symbol $\mathcal{F}_X$. We have $\mathcal{F}_X \subset \mathcal{F}$ and a measurement of $X$ completely determines $\omega \in \Omega$ if the random variable $X$ is one-to-one or when $\mathcal{F}_X = \mathcal{F}$. Moreover, finer sub-$\sigma$-algebra measurements provide better search results for $\omega \in \Omega$.

Alfr{\'e}d R{\'e}nyi describes the information content in a random variable in terms of the $\sigma$-algebra it generates \cite{renyi2007probability} \cite{rota2001twelve}. We see this perspective in standard probability textbooks \cite{durrett2019probability,van2007stochastic}, particularly in the chapters of conditional expectation and martingales---where a time-indexed filtration represents all information known up until that time. 

Consider $ \mathcal{G} \subset \mathcal{F}_X$, a sub-$\sigma$-algebra that provides partial information about $X$. Note that $X$ is not $\mathcal{G}$-measurable, but Kolmogorov's probability theory has a natural way of finding a $\mathcal{G}$-measurable random variable such that it is as close as possible to $X$ by means of a probability measure $\vp$ on $\mathcal{F}$. This random variable is called the conditional expectation $\mathbb{E}\left[ X \mid \mathcal{G}\right]$ and serves as the ``information projection'' of $X$ onto the space of $\mathcal{G}$-measurable random variables. 
\begin{align}
\label{eq:infoproj-prob}
        \mathbb{E}\left[ X \mid \mathcal{G}\right] = \operatorname*{arg\,inf}_{\mathcal{G}-\text{measurable } Z } \mathbb{E}\left[ (X-Z)^2 \right].
\end{align}

This orthogonal projection occurs in the Hilbert space $L^2( \Omega, \mathcal{F}, \vp)$ of square-integrable random variables, where the inner product $\langle X,Y \rangle = \mathbb{E} [XY]$. This notion of projection is a familiar concept from elementary geometry and linear algebra, where projecting a point onto a subspace means finding the closest point by means of a distance in a subspace.

The present work investigates the equivalence in the notion of information projection from IG to the natural idea of information projection using probability theory. This idea is emphasized in Theorem \ref{lemma:infoproj}.

%%%%%%%%%%%%%%%%%% METHODS AND RESULTS %%%%%%%%%%%%%%%%%%
    \section{Methods and Results}
\label{sec:3}

As we have alluded to, the assumptions on the nature of repeated samples are more basic than the assumptions on the probability. In this section, we first restrict our discussion to the category of i.i.d. samples in a sequence, followed by considering sample sequences with a Markov dependency but identical distribution.

\subsection{Independent and identically distributed samples category}
\label{sec:3_iid}

We consider a series of $N$ repeated experiments with data collected from each iteration. We treat each repetition as identical to facilitate an identical probability measure for each experiment. We draw this assumption from non-linear dynamical systems, specifically ergodic systems, where we see the notion of shift-invariance.

 Consider the finite state space $\mathcal{S} = \{1, 2, \ldots, n\}$, and conduct an experiment where we sample from this set $N$ times with replacement. Even with the assumption of identical trials, one can define numerous measures on the discrete sigma-algebra generated by $\Omega=\mathcal{S}^N$, namely $2^{\Omega}=2^{\mathcal{S}^N}$. For example, assuming independent samples and Markovian dependencies give different probability measures. Moreover, in our scenario, we focus on a fixed number $N$ of samples and take $N \rightarrow \infty$. For infinite sequences, the larger space of outcomes, $\Omega$, consists of all functions from $\mathbb{Z}_{>0}$ to $\Omega$, which is an uncountably infinite set. In this case, we can define numerous measures on the sigma-algebra generated by the larger $\Omega$ space.

Therefore, the assumption that the nature of repeated samples is identical is more fundamental than the assumptions about probability. In the present work, we first restrict our discussion exclusively to independent samples in a sequence so, to define our measure, we only need to know probabilities $\vp = (p_1,\ldots,p_n)$ we assign for each state. Then, in Sec. \ref{sec:3_markov}, we assume a stationary Markov process for which one needs to know the transition probabilities; there is a conditional i.i.d. within this type of data.

 This approach parallels the experimental setup in quantum mechanics, where researchers often examine measurements across multiple quantum particles. These quantum systems may include particles that interact or do not interact. For example, photons do not interact with each other on their own unless we mediate the interaction \cite{Nielsen_Chuang_2010}. Double-slit and polarization experiments with monochromatic light waves are examples where experimenters make many measurements of several non-interacting photons. Analyzing multiple non-interacting, identical quantum particles is similar to i.i.d. samples in probability theory \cite{baym1969}.
 
 Let's assign probabilities $\vp = (p_1,\ldots,p_n)$ for each state. 
 In the limit of data {\em ad infinitum}, the Kullback-Leibler divergence (KLD)  $H(\vnu \mid \vp) =\sum_i^n\nu_i \log \frac{\nu_i}{p_i} $ is the rate function in Sanov's Large deviation theorem \cite{dembo-book}. 

\begin{example}
\label{example:infogain2}
    Continuing example \ref{example:infogain1}, in presence of a probability measure provided by i.i.d. probabilities $\vp = (p_1,\ldots,p_n)$, the information gain from knowing the empirical frequencies $\vnu \in \rd \vq$ relative to the entire space of sequences $\mathcal{S}^N$ is

    \[
    \log \frac{1}{\mathbb{P}[\vnu \in \rd \vq]} \approx   N H(\vq \mid \vp).
    \]
\end{example}
 
With $\vp$ as a fixed parameter, $H \colon \riDelta{n} \to \reals_{\geq 0} $ is called entropy, and it is a convex function of empirical frequencies. The Legendre-Fenchel transform of entropy $H(\,\cdot \mid \vp)$, free energy $F \colon \reals^n \to \reals$ is a function of energies $\vmu$ defined as
\begin{align*}
    F(\vmu \mid \vp) = \log \left( \sum_{i = 1}^n p_i e^{\mu_i} \right) =\log  \mathbb{E} \left[  e^{ \vmu } \right]. 
\end{align*}

Here, $\log  \mathbb{E} \left[  e^{ \vmu } \right]$ is also the cumulant generating function of the random variable $\mI_n$.
The  energies $\vmu$ are conjugate of empirical frequencies $\vnu$ and are related as follows
\begin{align*}
    \nu_i =\frac{\partial}{\partial \mu_i} F(\vmu \mid \vp)=  \frac{p_i  e^{\mu_i}}{ \left( \sum_{j = 1}^n p_j e^{\mu_j} \right)} = p_i e^{ \mu_i - F(\vmu \mid \vp)}.
\end{align*}

The function $F(\vmu \mid \vp)$ is the cumulant generating function (CGF) of the indicator random variables $\mI_n$ introduced in Example~\ref{ex:indicator}, with $\vmu$ serving as a parameter. Moreover, for a large $N$-trial, the function $N F(\vmu \mid \vp)$ is the CGF of the random variable $N \vnu$, the unnormalized counting frequency. 

Consider the unnormalized count of each symbol, $\vc = N \vnu$, a random variable on the sequence space $\mathcal{S}^N$ as opposed to the singleton event space $\mathcal{S}$. We write the cumulant generating function $ \Lambda^{\vc}(\vmu \mid \vp)$ as

\begin{align*}
    \Lambda^{\vc}(\vmu \mid \vp) = \log \mathbb{E}\!\left[e^{\vmu^T \vc}\right] = \log \mathbb{E}\!\left[e^{N\,\vmu^T \vnu}\right].
\end{align*}

We start from
\begin{align*}
    e^{\Lambda^{\vc}(\vmu \mid \vp)}
    = \mathbb{E}\!\left[e^{\vmu^T \vc}\right].
\end{align*}
Write the expectation as a sum over all sequences (paths) 
$\omega \in \mathcal{S}^N$ with $\mathcal{S} = \{1,\dots,n\}$.  
Let $\vc(\omega) = (c_1(\omega),\dots,c_n(\omega))^T$ be the count vector for $\omega$.  
Under the i.i.d.\ base law $\vp$,
\[
    \Prob(\omega) = \prod_{i=1}^n p_i^{c_i(\omega)}.
\]
Then
\begin{align*}
    e^{\Lambda^{\vc}(\vmu \mid \vp)}
    = \sum_{\omega \in \mathcal{S}^N}
       e^{\vmu^T \vc(\omega)} \prod_{i=1}^n p_i^{c_i(\omega)} = \sum_{\omega \in \mathcal{S}^N}
       \prod_{i=1}^n \bigl(p_i e^{\mu_i}\bigr)^{c_i(\omega)}.
\end{align*}

Multiply and divide each term in the sum by $e^{N F(\vmu \mid \vp)}$:
\begin{align*}
    \prod_{i=1}^n \bigl(p_i e^{\mu_i}\bigr)^{c_i(\omega)}
    &= e^{N F(\vmu \mid \vp)}
       \prod_{i=1}^n
       \Bigl(p_i e^{\mu_i - F(\vmu \mid \vp)}\Bigr)^{c_i(\omega)}.
\end{align*}
The factors inside the product define the tilted probability or a single sample
\[
    p_i^{\vmu} = p_i e^{\mu_i - F(\vmu \mid \vp)}, \quad i=1,\dots,n,
\]
so that for each path, the tilted probability measure is
\[
    \prod_{i=1}^n
    \Bigl(p_i e^{\mu_i - F(\vmu \mid \vp)}\Bigr)^{c_i(\omega)}
    = \Prob^{\vmu}(\omega),
\]
the probability of $\omega$ under the tilted product measure. Hence
\begin{align*}
    e^{\Lambda^{\vc}(\vmu \mid \vp)}
    &= e^{N F(\vmu \mid \vp)}
       \sum_{\omega \in \mathcal{S}^N} \Prob^{\vmu}(\omega)
     = e^{N F(\vmu \mid \vp)} \,(1)
     = e^{N F(\vmu \mid \vp)}.
\end{align*}
Taking logarithms gives
\[
    \Lambda^{\vc}(\vmu \mid \vp) = N F(\vmu \mid \vp).
\]

This leads to a relationship between the Hessian of $F$ and the Covariance of $\vc$ under a $\vmu$-tilted measure.
\begin{align}
\label{eq:hessian-cov1}
     \frac{1}{N}\,
    \Lambda^{\vc}(\vmu \mid \mP)
    &= F(\vmu \mid \mP), \notag\\
     \frac{1}{N}\,
    \nabla_{\vmu} \Lambda^{\vc}(\vmu \mid \mP)
    &=  \nabla_{\vmu} F(\vmu \mid \mP) = \mathbb{E}^{\vmu}\!\left[\vnu\right], \notag\\
    \frac{1}{N}\,
    \nabla_{\vmu}^{\,2} \Lambda^{\vc}(\vmu \mid \mP)
    &= \nabla_{\vmu}^{\,2} F(\vmu \mid \mP)\notag\\
    &= \frac{1}{N}
       \left(
          \mathbb{E}^{\vmu}\left[\vc\left(\vc\right)^{T} \right]
          - \mathbb{E}^{\vmu}\!\big[\vc\big]\,
            \mathbb{E}^{\vmu}\!\big[\vc\big]^{T}
       \right)\notag\\
    &= \mathrm{Cov}^{\vmu}\!\left[\sqrt{N}\,\vnu\right].
\end{align}

\begin{notation} 
    We denote the measure $p_i e^{\mu_i - F(\vmu \mid \vp)}$ for $i \in \mathcal{S}$ as $\vp^{\vmu}$ and call it the exponentially tilted measure, tilted by $\vmu$.

    We borrow this terminology from importance sampling \cite{Siegmund-importance-sampling}. For $n \in \mathbb{Z}_{>0}$, a scalar $c \in \reals$ and vector $\vy = \bmat{y_1 & y_2 &\ldots &y_n}^T \in \reals^n$, we define the operation $\vy + c \coloneq \vy + c\ones_n = \bmat{y_1 + c & y_2 + c &\ldots &y_n +c}^T $. We define $e^{\vy} = \exp{\vy} \coloneq \bmat{e^{y_1} & e^{y_2} &\ldots &e^{y_n}}^T $. For $\vy \in \reals^n$ and $\mathcal{C} \subset \reals^n$, define $\vy + \mathcal{C}\coloneq \left\{ \vx + \vy \mid  \vx \in \mathcal{C} \right\}$.
\end{notation}

Let $e^{ \vmu - F(\vmu \mid \vp)} $ be a random variable on $\mathcal{S}$ taking values $ e^{ \mu_i - F(\vmu \mid \vp)}$ for $i \in \mathcal{S}$. The random variable $e^{ \vmu - F(\vmu \mid \vp)} $ is a Radon-Nikodym derivative because it is positive and has expectation
\begin{align*}
    \mathbb{E} \left[  e^{ \vmu - F(\vmu \mid \vp)} \right] = \sum_{i=1}^n  p_i e^{ \mu_i - F(\vmu \mid \vp)} = \sum_{i=1}^n \nu_i =1.
\end{align*}

Notice that the space of empirical frequencies, $\riDelta{n}$ is one dimension less than that of energies $\reals^n$; this is because empirical frequencies depend on each other via the relation $\sum_{i=1}^n \nu_i = 1$. A single $\vnu \in \riDelta{n}$ has infinitely many conjugates $\vmu$. That is, if $\vmu$ is a conjugate of $\vnu$ then, $\vmu + c$ is also a conjugate of $\vnu$. This follows a similar principle in science: energy is not an absolute quantity but is always measured relative to a reference point.

\begin{lemma}{(Free Energy Chain Rule)}\label{lemma:free-energy-chain}
    Let $\ones_n$ denote a $n$-vector of ones. For any $\vmu^{(1)}, \vmu^{(2)} \in \reals^n$, we have 
    \begin{align*}
        F(\vmu^{(1)} + \vmu^{(2)} \mid \vp) = F\left(\vmu^{(1)}\mid \vp\right) +  F\left(\vmu^{(2)} \mid \vp^{\vmu^{(1)}} \right) 
    \end{align*}

    \begin{proof}
    
    Starting with the definition of free energy of $\vmu^{(1)}+ \vmu^{(2)}$
    \begin{align*}
    F(\vmu^{(1)} + \vmu^{(2)} \mid \vp) &= \log \left( \sum_{i = 1}^n p_i e^{\mu^{(1)}_i + \mu^{(2)}_i} \right) = \log \left( \sum_{i = 1}^n p_i e^{\mu^{(1)}_i - F(\vmu^{(1)}\mid \vp) }e^{ \mu^{(2)}_i + F(\vmu^{(1)}\mid \vp)} \right) \\
    &= F\left(\left. \vmu^{(2)} + \ones_n F\left(\vmu^{(1)}\mid \vp\right) \, \right| \vp^{\vmu^{(1)}} \right) = F\left(\vmu^{(1)}\mid \vp\right) +  F\left(\vmu^{(2)} \mid \vp^{\vmu^{(1)}} \right) 
\end{align*}
    \end{proof}
\end{lemma}

We see that the change in reference point for energies corresponds to the exact change in reference for free energy in the following corollary.
\begin{corollary}
\label{cor:free-energy-degen}
For a scalar $c \in \reals$, $F(\vmu + c \mid \vp) = c + F(\vmu \mid \vp). $

\begin{proof}
 This follows directly from the fact that $\vp^{c\ones_n} = \vp$ and Lemma \ref{lemma:free-energy-chain}. 
\end{proof}
\end{corollary}

The previous corollary should help us realize that the free energy function is not strictly convex since $F\left(\frac{1}{2}\left((\vmu + 2c) + (\vmu)\right) \mid \vp\right) = \frac{1}{2}\left(2c + F(\vmu \mid \vp)\right) + \frac{1}{2} F(\vmu \mid \vp). $

The Hessian of $H$ provides the metric on $ \riDelta{n} \subset\mathbb{R}^n$, the space of $\vnu$; 
and the Hessian of $F$ provides the metric on  $\reals^n$, the space of $\vmu$.
\begin{align*}
    g^{ij}(\vmu) &= \frac{\partial^2}{\partial \mu_i \partial \mu_j} F(\vmu \mid \vp) = \frac{\partial}{\partial \mu_j} \nu_i = \frac{\partial}{\partial \mu_j} p_i e^{ \mu_i - F(\vmu \mid \vp)} \\
    &=  \delta_i^j \left(p_ie^{\mu_i - F(\vmu \mid \vp)}\right) - \left(p_ie^{\mu_i - F(\vmu \mid \vp)}\right)\left(p_je^{\mu_j - F(\vmu \mid \vp)}\right) \\
    &= \left(p_ie^{\mu_i - F(\vmu \mid \vp)}\right)\left( \delta_i^j - p_je^{\mu_j - F(\vmu \mid \vp)} \right).
\end{align*}
We notice that $g(\vmu)$ is the covariance matrix of the random variable $\mI_n$ with the probability distribution $\vp^{\vmu}$ and moreover, for an arbitrary random variable $\vy \colon \mathcal{S} \to \reals$, the inner-product of $\frac{\partial}{\partial \mu_i} F(\vmu \mid \vp)$ and $\vy$ is 
\begin{align*}
    \vy^T \frac{\partial}{\partial \mu_i} F(\vmu \mid \vp) = \vy^T \vnu = \mathbb{E}^{\vnu}\left[\vy\right].
\end{align*}
Since $g(\vmu)$ provides the metric in the space of $\vmu$, the space of all random variables in $\Omega$, for random variables $\vy \colon \mathcal{S} \to \reals$ and $\vz \colon \mathcal{S} \to \reals$  we can compute the bilinear form,

\begin{align*}
    \vy^T g(\vmu) \vz &= \sum_{i=1}^n \sum_{j=1}^n g^{ij}(\vmu)y_iz_j \\
    &= \sum_{i=1}^n \sum_{j=1}^n \left(p_ie^{\mu_i - F(\vmu \mid \vp)}\right)\left( \delta_i^j - p_je^{\mu_j - F(\vmu \mid \vp)} \right)y_iz_j \\
    &= \sum_{i=1}^n y_iz_i p_ie^{\mu_i - F(\vmu \mid \vp)} - \left( \sum_{i=1}^n y_i p_ie^{\mu_i - F(\vmu \mid \vp)} \right) \left( \sum_{i=1}^n z_i p_ie^{\mu_i - F(\vmu \mid \vp)} \right)\\
    &= \mathbb{E}^{\vnu}\left[\vy\vz\right] - \mathbb{E}^{\vnu}\left[\vy\right] \mathbb{E}^{\vnu}\left[\vz\right].
\end{align*}
The bilinear form results in the covariance between random variables $\vy$ and $\vz$ under the empirical frequency $\vnu$ as a probability measure. Suppose that $\vz = \vy$, then we obtain the quadratic form

\begin{align*}
    \vy^T g(\vmu) \vy = \mathbb{E}^{\vnu}\left[\vy^2\right] - \left(\mathbb{E}^{\vnu}\left[\vy\right]\right)^2.
\end{align*}

The quadratic form results in the random variable $\vy$ variance under the empirical frequency $\vnu$ as a probability measure.

\subsection{Contraction of empirical frequencies}
\label{subsection:contraction}
Now consider a random variable $\vX \colon \mathcal{S} \to \reals^k$, taking values $\vx_1, \ldots, \vx_n \in \reals^k$.  After $N$ observations ${\vX}^{(1)}, \ldots, {\vX}^{(N)}$, we compute the empirical mean as:

\[
\bar{\vx}  = \frac{1}{N} \sum_{j=1}^N {\vX}^{(j)}.
\]
 We express this through empirical frequencies as $\vx = \sum_{i=1}^n \nu_i \vx_i$. Using the matrix-form notation from notation \ref{rv-matrix}, the random variable $\vX$ has $\vx_1, \ldots, \vx_n$ as columns. It follows that

\begin{align*}
    \vX = \begin{bmatrix}
    | & | & & | \\
    \vx_1 & \vx_2 & \cdots & \vx_n \\
    | & | & & |
\end{bmatrix} \text{ and,  }\vx = \vX\vnu.
\end{align*}

Throughout our discussion, we will assume that the rowspace of $\vX$ is full-rank, does not contain $\bmat{1 & 1 \ldots & 1}$, and $k<n-1$. Suppose $k=n-1$, the rows of $\vX$ are linearly independent, and the rowspace does not contain $\bmat{1 & 1 \ldots & 1}$. In that case, the random variable $\vX$  provides {\em holographic} information about $\vnu$ via $n$ linearly independent equations, and we may define a change of basis from the space of $\vnu$ to the space of $\vx$. 
The empirical frequencies for the empirical mean $\vx$ fall in the space $\vnu \in \mathcal{A}_\vx = \{\vnu \in \reals^n_{> 0}\mid \sum_{i=1}^{n}\nu_i = 1, \sum_{i=1}^n \nu_i \vx_i = \vx \}$.

Applying the contraction principle \cite{dembo-book} to empirical means $\vx$, we obtain
\[
    \log \Prob[\vx \in \rd \vy]
    \approx   -N \inf_{\vq \in \mathcal{A}_{\vy}} H(\vq \mid \vp).
\]
We therefore identify the rate function for the empirical mean as
\begin{align}
\label{eq:entropy-contraction}
    \phi(\vx \mid \vp)
    &= \inf_{\vq \in \riDelta{n}}
       \left\{
           H(\vq \mid \vp)
           \,\bigg|\,
           \sum_{i=1}^n q_i \vx_i = \vx
       \right\} \\
    &= \inf_{\vq \in \riDelta{n}}
       \left\{
           \left.
           \sum_{i=1}^n q_i \log\!\left(\frac{q_i}{p_i}\right)
           \,\right|\,
           \sum_{i=1}^n q_i \vx_i = \vx
       \right\}. \notag
\end{align}

In convex analysis \cite{rockafellar-book}, the authors typically denote contractions involving linear transformations as $\phi = (\vX H)$.
The rate function $ \phi(\vx \mid \vp) $ is a convex function of $\vx$ \cite{rockafellar-book}, and it is the entropy function as a function of empirical mean $\vx$ and the probability measure $\vp$. The domain of $\vx$ is the \textit{open convex hull} of $\vx_1,\ldots, \vx_n$ which we denote as $\Conv{\vX}  = \Conv{\vx_1,\ldots, \vx_n }$.

\begin{example}
\label{example:infogain3}
    Continuing example \ref{example:infogain2}, in presence of a probability measure provided by i.i.d. probabilities $\vp = (p_1,\ldots,p_n)$, the information gain from knowing the empirical mean $\vx$ relative to the entire space of sequences $\mathcal{S}^N$ is

    \[
    \log \frac1{\mathbb{P}[\vx \in \rd \vy]} \approx   N \phi(\vx \mid \vp).
    \]

    This gives an information-theoretic meaning to the large deviation rate function $\phi$. 
    
\end{example}

Following dual operations (see p. 142, Theorem 16.3 of \cite{rockafellar-book}), we simplify the rate function as
\begin{align*}
  \phi(\vx \mid \vp)  &= \inf_{\vy }\left\{ \vy^T \vx - \log \left( \sum_{i=1}^n p_i e^{\vy^T \vx_i}\right)\right\} \\
  &=\valpha^T \vx - \log \left( \sum_{i=1}^n p_i e^{\valpha^T \vx_i}\right)
\end{align*}
Where $\valpha \in \reals^{k}$ is the unique solution to $\sum_i p_i e^{\valpha^T \vx_i} \vx_i = \vx \sum_{i=1}^n p_i e^{\valpha^T \vx_i}$.  This follows from the fact that $\log \left( \sum_{i=1}^n p_i e^{\valpha^T \vx_i}\right)$ is a smooth convex function of $\valpha \in \reals^k$ and no two values of $\alpha$ can have the same gradient. This defines a diffeomorphism between $\Conv{\vX} \subset \reals^k$, the space of $\vx$ and $\reals^k$, the space of $\valpha$. The duality also establishes that 

\[\nu^*_j = \frac{p_j e^{\valpha^T \vx_j}}{\sum_{i=1}^n p_i e^{\valpha^T \vx_i}}  , \text{ for } j \in \mathcal{S} \]

is the optimizer for the optimization problem in equation \eqref{eq:entropy-contraction}, in the form of an exponential family of distributions with parameters $\valpha$ in the discrete space. Alternatively, one can also say that the empirical means $\vx \in \Conv{\vX}$ determine an embedding within $\riDelta{n}$ via the relation $\vnu_{\vx} = \vp^{\vX^T \nabla \phi(\vx \mid\vp) }$. Exponential family of distributions are thoroughly explored in previous works of IG \cite{amari-book} \cite{JMLR:v6:banerjee05b}, statistical mechanics \cite{GibbsBook}, thermodynamics including the works like those of Szilard \cite{szilard}. This family includes some of the most frequently used statistical models, including Gaussian, Gamma, Poisson, and Geometrical.

A reader familiar with the Legendre-Fenchel transform can identify $\valpha$ as the conjugate variable. Fixing our probabilistic measure $\vp$, the Legendre-Fenchel transform of our entropy function $\phi(\vx \mid \vp)$ is 
\begin{align*}
    \psi (\valpha \mid \vp) &= \sup_{\vx} \left\{ \valpha^T \vx - \phi(\vx \mid \vp)  \right\} \\
    &=\log \left( \sum_{i=1}^n p_i e^{\valpha^T \vx_i}\right).
\end{align*}

The function $\psi (\valpha \mid \vp) $ is the cumulant generating function (CGF) of the random variable $\vX$ with parameter $\valpha$. 

As a consequence of Legendre-Fenchel transform, we notice that $ \psi $ is the lift of $F$, that is,
\begin{align}
     \psi (\valpha \mid \vp)  = F (\valpha^T X \mid \vp). \label{eq:free-energy-contraction}
\end{align}
In convex analysis \cite{rockafellar-book}, the authors typically denote lift involving linear transformations as $\psi = (F \vX )$.

The Hessian of $\phi$ and $\psi$ provide the metric on the space of $\vx$, $\Conv{\vX}$ and the dual space of $\valpha$, $\reals^k$ respectfully. The Legendre-Fenchel transform provides a framework within which the Hessians of dual functions are inverses of each other. When $\vx$ and $\valpha$ are Legendre conjugates of each other, then $\vx = \nabla_\valpha \psi(\valpha) $ and  $\valpha = \nabla_\vx \phi(\vx) $. We have 
\begin{align*}
    \nabla_\valpha\nabla_\valpha  \psi(\valpha) = \left( \nabla_\vx \nabla_\vx \phi(\vx) \right)^{-1}.
\end{align*}

Using the contraction from equation \eqref{eq:free-energy-contraction}, the Hessians of $\psi$ and $F$ are related as

\begin{align}\label{eq:alpha-metric-transf}
    \nabla_\valpha\nabla_\valpha  \psi(\valpha) = \vX^T \left(\nabla_{\vmu} \nabla_{\vmu} \left.F(\vmu)\right|_{\vmu = \alpha^T \vX} \right) \vX.
\end{align}

The linear transformation $\vmu = \vX^T \valpha$ embeds the space of $\valpha \in \reals^k$ within the space of $\vmu \in \reals^n$. Riemannian geometry \cite{Lee2002-gr} naturally transforms a metric to an embedding, aligning with equation \eqref{eq:alpha-metric-transf}.

The Hessian of $\psi$ simplifies to 
\begin{align}
\label{eq:variance-hessian}
    \nabla_\valpha\nabla_\valpha  \psi(\valpha) = \text{Cov}^{\vp^{\vX^T \valpha}}\left(\vX \right).
\end{align}
The Hessian of $\psi$ results in the covariance of the random variable $\vX$ under  $\vp^{\vX^T\valpha}$ as a probability measure. Equation \eqref{eq:variance-hessian} is well known in exponential family theory  \cite{dasgupta2011probability}. We re-deduce it here to characterize the geometry of the space of $\vx$. Typically, one calculates the empirical covariance directly from the observed data or using the empirical frequency. However, in our context, we focus on the space of empirical means $\vx$ and its conjugate variable $\valpha$, rather than the empirical frequencies. The covariance emerged from the Hessian of $\psi$ without knowing the empirical frequencies $\vnu$, indirectly from the data. Hence, we will call this quantity the ``explained covariance''.   If the empirical mean $\vx$ uniquely determines the empirical frequency $\vnu$ through the set of linear equations $\sum_{i=1}^n \nu_i \vx_i = \vx$, we call this scenario \textit{holographic}. The information provided by the random variable $\vX$ is complete. If the equation admits infinitely many solutions, then this reflects incomplete information. This perspective reflects E. T. Jaynes' view  ~\cite{JaynesBook} that for general problems of scientific inference, almost all arise from incomplete information rather than `randomness’. Information geometry represents empirical data {\em ad infinitum}, which means that there is no randomness. When information is incomplete, we view probability theory as an extension of logic that fills in when deductive reasoning falls short. Within our framework, we use the theory of large deviation and the contraction principle to characterize the explained covariance and represent it using the Hessian of $\psi$.

In the space of $\vx$, the Riemannian metric is the inverse of the explained variance of the random variable $\vX$. 
\begin{align}
    \nabla_\vx \nabla_\vx \phi(\vx) = \text{Cov}^{\vnu}\left(\vX \right)^{-1} \label{metric-hessian-cov}
\end{align}

This further deepens the implication of using the Hessian of the rate function as a Riemannian metric in the space of $\vx$. Each random variable in $\vX$ may represent a different measured quantity with different scales and units, and the standard Euclidean metric is fundamentally inadequate and inappropriate for any analysis. A metric that accounts for variability and correlation among measurements is required; this is achieved by the inverse of the covariance matrix. This form of rescaling is analogous to non-dimensionalization in differential equations and physics. In scientific and engineering settings, decisions about units and scaling are often driven by measurement error, which strengthens our rationale for the metric.

It is useful to note a qualitative difference in the geometry of empirical frequencies $\vnu$ and the geometry of empirical means $vx$. The metric in the space of empirical frequencies does not depend on the underlying i.i.d measure $\vp$. Indeed, because we can compute the covariance directly from $\vnu$ without reference to $\vp$. In this sense, under the assumption of i.i.d, the geometry of empirical frequencies is measure independent. 

By contrast, when the mapping $\vx = \vX \vnu$ is not injective, the empirical mean alone cannot determine the second-order statistics. In such cases, the underlying measure $\vp$ helps in inferring missing information. The geometry of empirical mean data depends on the underlying measure $\vp$.

\begin{example}
    Consider two independent unit-variance random variables $X_1$ and $X_2$. Since the random variables possess unit variance and are independent, the Euclidean metric makes sense for measuring the distance between measurements. That is, the distance element $\Delta s$ is found using

    \begin{align*}
        (\Delta s)^2 = (\Delta x_1)^2 + (\Delta x_2)^2.
    \end{align*}
    
    The metric with the coordinates $(x_1,x_2)$ is $g_{ij} = \delta_i^j$.
    We now introduce a covariance factor $0<\rho<1$, consider two correlated, unit-variance random variables $Y_1= X_1,Y_2= \rho X_1 + \sqrt{1 - \rho^2} X_2$ which is written as
    \begin{align*}
       \vY= \begin{bmatrix}
     Y_1 \\
     Y_2 
    \end{bmatrix}
    = 
    \begin{bmatrix}
    1 & 0 \\
    \rho & \sqrt{1-\rho^2}
    \end{bmatrix}
    \begin{bmatrix}
    X_1 \\
    X_2
    \end{bmatrix} = L \vX,
    \end{align*}
    
    where $L = \begin{bmatrix}
    1 & 0 \\
    \rho & \sqrt{1-\rho^2}
    \end{bmatrix}.$
    
    The covariance between $Y_1$ and $Y_2$ is 
    
    \begin{align*}
        \mathbb{E}[Y_1Y_2] - \mathbb{E}[Y_1]\mathbb{E}[Y_2] =  \rho\left(\mathbb{E}[X_1^2] - \mathbb{E}[X_1]^2\right) +   \sqrt{1-\rho^2} \left(\mathbb{E}[X_1X_2] - \mathbb{E}[X_1]\mathbb{E}[X_2]\right) = \rho.
    \end{align*}
    
    Since the transformation is linear, invertible, and non-identity, the new metric tensor representation $\Tilde{g}_{ij}$ is not $\delta_i^j$.
    Building on one of the fundamental principles of geometry, where distance is geometric object invariant under change of coordinates chart, we compute the distance element $\Delta s$ in terms of coordinates $(y_1,y_2)$.
    
    \begin{align*}
        (\Delta s)^2 &= (\Delta \vx)^T(\Delta \vx) = (L^{-1}\Delta \vy)^T (L^{-1}\Delta \vy)=(\Delta \vy)^T(LL^T)^{-1} (\Delta \vy) \\
        & = \begin{bmatrix}
     \Delta y_1 &
     \Delta y_2 
    \end{bmatrix}  \begin{bmatrix}
    1 & \rho \\
    \rho & 1
    \end{bmatrix}^{-1} \begin{bmatrix}
     \Delta y_1 \\
     \Delta y_2 
    \end{bmatrix}.
    \end{align*}
\end{example}

This example illustrates how the metric representation changes to the inverse of covariance between axes. Moreover, it is well-known for a multivariate Gaussian distribution $(\vmu,\Sigma)$ that the quadratic form $\frac{1}{2} (\vX-\vmu)^T \Sigma^{-1} (\vX-\vmu) $ in the exponent emerges naturally. This form mirrors the expression for the squared distance element $(\Delta s)^2$ in terms of the new coordinates $(y_1, y_2)$, illustrating the relationship between statistical concepts and geometric interpretations.

%%%%%%%%%%%%%%%%%%%%%%%%%%%
\subsection{Information Projection}
\label{subsection:info-proj}

We now show that the information projection, using the methods of probability
theory, unified with the notion of information projection in the IG
literature \cite{csiszar1975divergence}.

Recall that $\Omega = \mathcal{S}^N$ and $\mathcal{F} = 2^\Omega$. For each
$\omega \in \Omega$, the empirical frequency $\vnu(\omega)$ and empirical mean
$\vx(\omega) = \vX\vnu(\omega)$ are random variables on $(\Omega, \mathcal{F})$.
The empirical frequency takes values in the $N$-lattice simplex
$\Delta^n_N = \{\vq \in \Delta^{n} : N q_i \in \mathbb{Z}_{\geq 0}\ \forall
i\}$, and generates the $\sigma$-algebra:

\begin{align*}
    \mathcal{F}_{\vnu}
    = \sigma\!\left(\left\{\vnu^{-1}(\vq) : \vq \in \Delta^n_N\right\}\right)
\end{align*}

where $\vnu^{-1}(\vq) = \{\omega \in \Omega \,:\, \vnu(\omega) = \vq\}$. The empirical mean $\vx$ takes values in the finite lattice
of achievable means:    

\begin{align*}
    \mathcal{X}_N = \{\vX\vq \,:\, \vq \in \Delta^n_N\} \subset \mathbb{R}^k
\end{align*}

and generates the $\sigma$-algebra

\begin{align*}
    \mathcal{F}_{\vx}
    = \sigma\!\left(\left\{\vx^{-1}(\vy) \,:\, \vy \in \mathcal{X}_N\right\}\right).
\end{align*}

For $\delta > 0$, define the strictly coarser $\sigma$-algebra:

\begin{align}
\label{eq:delta-thick-sigmaalgebra}
    \mathcal{F}^\delta_{\vx}
    = \sigma\!\left(\left\{
        \vx^{-1}\left(B_\delta(\vy) \cap \mathcal{X}_N\right) \,:\, \vy \in \mathcal{X}_N
    \right\}\right)
\end{align}

where $B_\delta(\vy) = \{\vz \in \mathbb{R}^k \,:\, \|\vz - \vy\|_2 < \delta\}$
is the open $\delta$-ball around $\vy$. 

\begin{align*}
    \mathcal{F}
    \supseteq \mathcal{F}_{\vnu}
    \supseteq \mathcal{F}_{\vx}
    \supseteq \mathcal{F}^\delta_{\vx}.
\end{align*}

The coarse $\mathcal{F}^\delta_{\vx}$ is necessary because the set
$\{\vnu \in \Delta^{n} : \vX\vnu = \vx\}$ is a proper affine subspace of
$\Delta^{n}$ with an empty interior in the ambient topology, hence its
induced measure is singular with respect to the simplex. The
$\delta$-thickening restores a non-empty interior, rendering the Gibbs
conditioning principle applicable. We further show with an example that the theorem does not hold without $\delta$-thickening, establishing that this regularization is not merely a technical convenience but a necessity.

\begin{theorem}[Information Projection]
    \label{lemma:infoproj}
    Fix an arbitrarily small $\delta > 0$ and let
$\mathcal{F}^\delta_{\vx}$ be a $\delta$-coarse sub-$\sigma$-algebra as defined in Equation~\eqref{eq:delta-thick-sigmaalgebra}. Suppose $\vp$ has full
support on $\mathcal{S}$. Then as
$N \to \infty$, the information projection of $\vnu$ onto
$\mathcal{F}^\delta_{\vx}$ is
    \begin{align*}
        \mathbb{E}[\vnu \mid \mathcal{F}^\delta_{\vx}](\omega)
        = \operatorname*{arg\,inf}_{\mathcal{F}^\delta_{\vx}\text{-measurable }
          \vz}\; \mathbb{E}\!\left[\|\vz - \vnu\|_2^2\right]
        \xrightarrow{N\to\infty}
        \operatorname*{arg\,inf}_{\vq \in \Delta^{n}}
        \left\{
            H(\vq \mid \vp)
            \;\Big|\;
            \|\vX\vq - \vx(\omega)\|_2 < \delta
        \right\}
    \end{align*}

\begin{proof}

    Recall from Definition~\ref{empirical-freq} that
    \[
        \vnu = \frac{1}{N} \sum_{j=1}^N \mI_n^{(j)}.
    \]
    By linearity of conditional expectation,
    \begin{align*}
        \mathbb{E}[\vnu \mid \mathcal{F}^\delta_{\vx}]
        &= \mathbb{E}\!\left[\left. \frac{1}{N} \sum_{j=1}^N \mI_n^{(j)}
          \,\right| \mathcal{F}^\delta_{\vx}\right]
        = \frac{1}{N} \sum_{j=1}^N
          \mathbb{E}\!\left[\mI_n^{(j)} \,\Big|\, \mathcal{F}^\delta_{\vx}\right] \\
        &= \frac{1}{N} \cdot N\,
          \mathbb{E}\!\left[\mI_n^{(1)} \,\Big|\, \mathcal{F}^\delta_{\vx}\right]
        = \mathbb{E}\!\left[\mI_n^{(1)} \,\Big|\, \mathcal{F}^\delta_{\vx}\right],
    \end{align*}
    where the $\mI_n^{(j)}$ are identically distributed by the symmetry of the
    i.i.d.\ sequence under permutation, even though the $\omega_j$ may not be independent under the $\mathcal{F}^\delta_{\vx}$ conditioning.
    Furthermore, the $\ell^{\text{th}}$ component of the above expectation is
    \begin{align*}
        \mathbb{E}[\vnu \mid \mathcal{F}^\delta_{\vx}]_{\ell}
        = \mathbb{E}\!\left[\mI_n^{(1)} \,\Big|\,
          \mathcal{F}^\delta_{\vx}\right]_{\ell}
        = \mathbb{P}\!\left[
            \bigl(\mI_n^{(1)}\bigr)_{\ell} = 1
            \,\Big|\, \mathcal{F}^\delta_{\vx}\right]
        = \mathbb{P}\!\left[
            \omega_1 = \ell \,\Big|\, \mathcal{F}^\delta_{\vx}\right].
    \end{align*}
    The Gibbs conditioning principle \cite[Thm.~3.3.3]{dembo-book} is
    applicable here because the set
    $\{\vnu \in \Delta^n : \vX\vnu \in B_\delta(\vx(\omega))\}$
    has non-empty interior in $\Delta^n$, being the preimage of an open ball
    under the continuous affine map $\vnu \mapsto \vX\vnu$. Moreover, since
    $H(\cdot \mid \vp)$ is strictly convex and the constraint set is convex,
    the infimum over the interior coincides with the infimum over its closure,
    so the conditions of \cite[Thm.~3.3.3]{dembo-book} are satisfied.
    Applying the principle, for large $N$:
    \begin{align*}
        \mathbb{E}[\vnu \mid \mathcal{F}^\delta_{\vx}]
        &= \Bigl\{
            \mathbb{P}\bigl[\omega_1 = 1 \,\big|\, \mathcal{F}^\delta_{\vx}\bigr],\,
            \mathbb{P}\bigl[\omega_1 = 2 \,\big|\, \mathcal{F}^\delta_{\vx}\bigr],\,
            \ldots,\,
            \mathbb{P}\bigl[\omega_1 = n \,\big|\, \mathcal{F}^\delta_{\vx}\bigr]
          \Bigr\} \\
        &= \left\{\mathbb{P}\!\left[\omega_1 = \ell \,\left|\,
            \left\|\frac{1}{N}\sum_{j=1}^N \vX^{(j)} - \vx(\omega)\right\|_2
            < \delta \right.\right]\right\}_{\ell=1}^n
    \end{align*}
    and therefore
    \begin{align*}
        \lim_{N\to\infty} \mathbb{E}[\vnu \mid \mathcal{F}^\delta_{\vx}]
        = \operatorname*{arg\,inf}_{\vq \in \Delta^{n}}
          \left\{ H(\vq \mid \vp)
          \;\Big|\; \|\vX\vq - \vx(\omega)\|_2 < \delta \right\}.
    \end{align*}
    This finishes the proof.
\end{proof}
\end{theorem}

\begin{remark}
Theorem~\ref{lemma:infoproj} identifies information projection as an $L^{2}$
projection. The conditional expectation
$\mathbb{E}[\vnu \mid \mathcal{F}^\delta_{\vx}]$ is the orthogonal projection
of the random vector $\vnu$ onto the $\delta$-coarse sub-$\sigma$-algebra
generated using the empirical mean $\vx$. This result departs from the standard
presentation in information geometry, where information projection is defined
a priori as a minimization of a dissimilarity measure, typically the
Kullback-Leibler divergence, and is explicitly distinguished from Euclidean
or $L^{2}$ projection \cite{amari-book,nielsen2018information,csiszar1975divergence}.
Here, the projection is $L^{2}$ by construction. The minimization of the KL
divergence arises from the Gibbs conditioning principle and large deviation
theory.
 
We emphasize that the $\delta$-thickening is not merely a technical device.
For a fixed $N$ and sufficiently small $\delta > 0$ — specifically, for
$\delta < \frac{1}{2}\min_{\vy \neq \vy' \in \mathcal{X}_N}\|\vy - \vy'\|_2$, the neighborhoods $B_\delta(\vy)$ around distinct points in $\mathcal{X}_N$ are
disjoint, and hence $\mathcal{F}^\delta_{\vx} = \mathcal{F}_{\vx}$. However,
this does \emph{not} imply that $\mathbb{E}[\vnu \mid \mathcal{F}_{\vx}]$
coincides with the entropy minimizer
$\operatorname*{arg\,inf}_{\vq \in \Delta^n}\{H(\vq \mid \vp) \mid \|\vX\vq -
\vx(\omega)\|_2 < \delta\}$. The order of limits is essential: one must
first take $N \to \infty$ with $\delta$ fixed, and only then $\delta \to 0$.
Exchanging these limits changes the answer entirely.
 
To see why $\mathbb{E}[\vnu \mid \mathcal{F}_{\vx}]$ is \emph{not} the
 minimizer of $H( \vq \,\mid \, \vp)$ under the constraint, $\vX \vq = \vx{\omega}$ for finite $N$, consider the following example. Let
$\mathcal{S} = \{1, 2, 3\}$ and the random variable $X: \mathcal{S} \to
\mathbb{R}$ take the irrational values
\begin{align*}
    X(1) = \sqrt{2}, \qquad X(2) = \sqrt{3}, \qquad X(3) = \sqrt{5}.
\end{align*}
Since $\sqrt{2}$, $\sqrt{3}$, and $\sqrt{5}$ are linearly independent over
$\mathbb{Q}$, the empirical mean
\begin{align*}
    \vx(\omega) = \nu_1\sqrt{2} + \nu_2\sqrt{3} + \nu_3\sqrt{5}
\end{align*}
uniquely determines $(\nu_1, \nu_2, \nu_3)$ for any finite sequence $\omega$.
That is, for any finite $N$, observing $\vx(\omega)$ is equivalent to
observing $\vnu(\omega)$ exactly, so $\mathcal{F}_{\vx} = \mathcal{F}_{\vnu}$
and
\begin{align*}
    \mathbb{E}[\vnu \mid \mathcal{F}_{\vx}](\omega) = \vnu(\omega),
\end{align*}
 independent of the underlying i.i.d.\ measure $\vp$. 
\end{remark}

In the following subsection, we discuss a non-i.i.d. case of statistical data under a Markovian measure. Similar to the i.i.d. case, we justify our choice of the metric, and in Theorem \ref{thm:infoproj-pairs}, we show that the information projection, using the methods of probability theory, coincides with the notion of information projection stated in the seminal work of Csiszár et al. \cite{csiszar1987conditional}

%%%%%%%%%%%%%%%%%%  %%%%%%%%%%%%%%%%%%
\subsection{Markov dependent and identically distributed sequences category}
\label{sec:3_markov}

In place of empirical frequency $\vnu=(\nu_1,\cdots,\nu_n)$ we now consider transition pairs $\vnu^{(2)}=(\nu_{ij})_{n\times n}$ from data {\em ad infinity}, where 
\[
   \nu_{ij} = \frac{1}{N}\sum_{\ell=1}^{N}
    \indicator_i^{(\ell-1)} \indicator_j^{(\ell)},
\]
with
\[
  \sum_{j=1}^n \nu_{kj}=
  \sum_{i=1}^n \nu_{ik} \, \text{ and }\, 
   \sum_{i,j=1}^n \nu_{ij} = 1.
\]
We define this space as follows.

\begin{definition}[Pairwise-frequency simplex]
Fix $n\in\mathbb{Z}_{>1}$. The \emph{pairwise-frequency simplex} is
\[
  \widetilde{\Delta}^{n}
  \;\coloneqq\;
  \Bigl\{\, \vnu^{(2)}=(\nu_{ij})_{i=1,j=1}^n \in \reals_{\geq 0}^{\,n\times n}
  \ \Big|\
  \sum_{i=1}^n\sum_{j=1}^n \nu_{ij}=1,\quad
  \sum_{j=1}^n \nu_{kj} = \sum_{i=1}^n \nu_{ik}\ \ \forall\,k\in \{1,\ldots,n \} \Bigr\}.
\]
\end{definition}

As the definition reveals, the empirical pair counting has a stationary marginal distribution: {\em shift-invariant} \cite{dembo-book}.
In place of KLD, one now has a large deviation rate function
\begin{equation}
       H^{(2)}\big(\vnu^{(2)} \,|\,\mP\big) = 
       \sum_{i,j=1}^n \nu_{ij}\log 
       \frac{\nu_{ij}}{\displaystyle 
       \sum_{k=1}^n \nu_{ik}\, P_{ij}}, 
\end{equation}
in which $\mP=(P_{ij})_{n\times n}$ is an ergodic Markov transition probability matrix with all $P_{ij}> 0$ and the sum over each row being $1$. 

The function $H^{(2)}\big(\vnu^{(2)}\mid\mP\big)$ is convex for $\vnu^{(2)}\in\riPairDelta{n}$.
For any $\alpha\in [0,1]$, 
\[
   H^{(2)}\Big(\alpha\vnu_a^{(2)}
   +(1-\alpha)\vnu_b^{(2)} \,\Big|\, \mP\Big) \le \alpha H^{(2)}\big(\vnu_a^{(2)} \,\big|\, \mP\big) + (1-\alpha) H^{(2)}\big(\vnu_b^{(2)} \,\big|\, \mP\big), 
\]
where $\vnu_a^{(2)}$, $\vnu_b^{(2)}\in\tilde{\Delta}^{n}$.  Its LFT
\begin{equation}
    F^{(2)}\big(\vmu^{(2)} \mid \mP \big) = \sup_{\vnu^{(2)}\in\riPairDelta{n}}\left\{ \sum_{i,j=1}^n \nu_{ij}u_{ij} -     H^{(2)}\big(\vnu^{(2)}\mid\mP \big)  \right\} = \log\lambda_{\text{max}}\Big(
      \mP e^{\vmu^{(2)}} \Big),
\label{F2}
\end{equation}
where $\mP e^{\vmu^{(2)}}$ denotes the $n\times n$ matrix with entries $P_{ij}e^{u_{ij}}$, and 
$\sum_{i,j=1}^n \nu_{ij}u_{ij}$ is known as the Frobenius inner product of two matrices $\vnu$ and $\vmu^{(2)}$. Readers may refer to Dembo and Zeitouni's Large Deviations Techniques and Applications \cite{dembo-book} for further details and proof.
Eq. (\ref{F2}) implies 
\begin{equation}
\label{paircounting-LFTpair}
   \nu_{ij} = \frac{\partial}{\partial u_{ij}}F^{(2)}\big(
      \vmu^{(2)}\mid\mP \big) = \lambda^{-1}_{\text{max}}\Big(\mP e^{\vmu^{(2)}} \Big)\frac{\partial}{\partial u_{ij}} \lambda_{\text{max}}\Big(
      \mP e^{\vmu^{(2)}} \Big) = \frac{ P_{ij}e^{u_{ij} }   v_iw_j }{ \lambda_{\text{max}} \big(\mP e^{\vmu^{(2)}} \big) } ,
\end{equation}

in which $\vv=(v_1,\cdots,v_n)$ and $\vw=(w_1,\cdots,w_n)$ are the left and right eigenvectors of matrix $\mP e^{\vmu^{(2)}}$ corresponding to the principal eigenvalue $\lambda_{\text{max}}$.  Therefore
\begin{equation}
    \sum_{i=1}^n \nu_{ik}
    = v_kw_k =  
    \sum_{j=1}^n \nu_{kj} 
     \text{ and }  
    \sum_{k=1}^n v_kw_k = 1.
\end{equation}

Define the $\vmu^{(2)}$-tilted transition kernel
\[
    P^{\vmu}_{ij}
    = \frac{P_{ij} e^{\mu^{(2)}_{ij}} w_j}{\lambda w_i},
\]
which satisfies $\sum_j P^{\vmu}_{ij} = 1$ for all $i \in \{1,\ldots,n\}$. The $\vmu^{(2)}$-tilted transition probability matrix is also ergodic with all $P^{\vmu}_{ij} > 0$. Therefore, the first moment of a cumulant generating function provides the stationary empirical pair-counting frequencies under the $\vmu^{(2)}$-tilted transition kernel.  

\paragraph*{Fenchel-Young inequality} 
For all $\vnu^{(2)}\in\riPairDelta{n}$ and all $\vmu^{(2)}\in\mathbb{R}^{n\times n}$, the Fenchel-Young inequality is as follows
\[
  H^{(2)}\big(\vnu^{(2)}\mid\mP\big) \;+\; F^{(2)}\big(\vmu^{(2)}\mid\mP\big)
  \;-\; \sum_{i,j=1}^n \nu_{ij}u_{ij} \ge 0,
\]
with equality if and only if $\vnu^{(2)}$ and $\vmu^{(2)}$ are LFT-pair, that is, when they satisfy the equality in equation \ref{paircounting-LFTpair}.

%%%%%%%%%%%%%%%%%%%%% CGF %%%%%%%%%%%%%%%%%%%%%%%%%%%%

\subsection{Cumulant Generating Function}
\label{cgf}

Consider a sequence of length $N$ consisting of symbols in $\{1,\ldots,n\}$. We examine the unnormalized count of pairs of symbols, $\vc^{(2)} = N \vnu^{(2)}$. We examine the cases where the underlying measure is a stationary Markov measure with transition probability $\mP$. We will use the cumulant generating function to establish a relationship between the Hessian of the free energy and the covariance of the statistical data.

We use $\vc^{(2)} = (c_{ij})_{i,j=1}^n$ to denote the pairwise count along a path 
\[
    \omega = (\omega_1,\omega_2,\dots,\omega_N) \in \mathcal{S}^{N+1},
    \qquad \mathcal{S} = \{1,\dots,n\}.
\]

The cumulant generating function of $\vc^{(2)}$ under a stationary Markov chain with transition matrix $\mP = (P_{ij})$ and stationary distribution $\vpi$ is
\[
    \Lambda^{\vc^{(2)}}(\vmu^{(2)} \mid \mP)
    = \log \mathbb{E}\!\left[\exp\bigl(\langle \vmu^{(2)}, \vc^{(2)}\rangle\bigr)\right],
\]
where $\langle \vmu^{(2)}, \vc^{(2)}\rangle = \sum_{i,j} \mu^{(2)}_{ij} c_{ij}$.

The probability of a single path $\omega$ under the stationary Markov measure is
\[
    \Prob(\omega)
    = \pi_{\omega_1} \prod_{k=0}^{N-1} P_{\omega_k \omega_{k+1}}
    = \pi_{\omega_1} \prod_{i,j} P_{ij}^{\,c_{ij}(\omega)}.
\]
Therefore
\begin{align*}
    e^{\Lambda^{\vc^{(2)}}(\vmu^{(2)} \mid \mP)}
    &= \mathbb{E}\!\left[\exp\bigl(\langle \vmu^{(2)}, \vc^{(2)}\rangle\bigr)\right] \\
    &= \sum_{\omega}
       \Prob(\omega)\,
       \exp\bigl(\langle \vmu^{(2)}, \vc^{(2)}(\omega)\rangle\bigr) \\
    &= \sum_{\omega}
       \pi_{\omega_1}
       \prod_{i,j} P_{ij}^{\,c_{ij}(\omega)}
       \exp\Bigl(\sum_{i,j} \mu^{(2)}_{ij} c_{ij}(\omega)\Bigr) \\
    &= \sum_{\omega}
       \pi_{\omega_1}
       \prod_{i,j} \bigl(P_{ij} e^{\mu^{(2)}_{ij}}\bigr)^{c_{ij}(\omega)}.
\end{align*}

The cumulant generating function is
\[
    \Lambda^{\vc^{(2)}}(\vmu^{(2)} \mid \mP)
    = \log \mathbb{E}\!\left[\exp\bigl(\langle \vmu^{(2)}, \vc^{(2)}\rangle\bigr)\right],
\quad
    \langle \vmu^{(2)}, \vc^{(2)}\rangle
    = \sum_{i,j} \mu^{(2)}_{ij} c_{ij}.
\]

Under the stationary Markov chain with transition matrix $\mP = (P_{ij})$ and stationary distribution $\vpi$, the path probability is
\[
    \Prob(\omega)
    = \pi_{\omega_1} \prod_{k=0}^{N-1} P_{\omega_k \omega_{k+1}}
    = \pi_{\omega_1} \prod_{i,j} P_{ij}^{\,c_{ij}(\omega)}.
\]
Thus
\begin{align*}
    e^{\Lambda^{\vc^{(2)}}(\vmu^{(2)} \mid \mP)}
    &= \mathbb{E}\!\left[\exp\bigl(\langle \vmu^{(2)}, \vc^{(2)}\rangle\bigr)\right] \\
    &= \sum_{\omega}
       \Prob(\omega)\,
       \exp\bigl(\langle \vmu^{(2)}, \vc^{(2)}(\omega)\rangle\bigr) \\
    &= \sum_{\omega}
       \pi_{\omega_1}
       \prod_{i,j} P_{ij}^{\,c_{ij}(\omega)}
       \exp\Bigl(\sum_{i,j} \mu^{(2)}_{ij} c_{ij}(\omega)\Bigr) \\
    &= \sum_{\omega}
       \pi_{\omega_1}
       \prod_{k=0}^{N-1}
       \Bigl(P_{\omega_k \omega_{k+1}} e^{\mu^{(2)}_{\omega_k \omega_{k+1}}}\Bigr).
\end{align*}

Recall the tilted transition kernel
\[
    P^{\vmu}_{ij}
    = \frac{P_{ij} e^{\mu^{(2)}_{ij}} w_j}{\lambda w_i}.
\]
 Inverting this relation gives
\[
    P_{ij} e^{\mu^{(2)}_{ij}}
    = \lambda P^{\vmu}_{ij} \frac{w_i}{w_j}.
\]
Substituting into the product, for each path $\omega$ we obtain
\begin{align*}
    \prod_{k=0}^{N-1}
    \Bigl(P_{\omega_k \omega_{k+1}} e^{\mu^{(2)}_{\omega_k \omega_{k+1}}}\Bigr)
    &= \prod_{k=0}^{N-1}
       \left(\lambda P^{\vmu}_{\omega_k \omega_{k+1}}
             \frac{w_{\omega_k}}{w_{\omega_{k+1}}}\right) \\
    &= \lambda^{N}
       \left(\prod_{k=0}^{N-1} P^{\vmu}_{\omega_k \omega_{k+1}}\right)
       \frac{w_{\omega_1}}{w_{\omega_N}}.
\end{align*}
Hence
\begin{align*}
    e^{\Lambda^{\vc^{(2)}}(\vmu^{(2)} \mid \mP)}
    &= \lambda^{N}
       \sum_{\omega}
       \pi_{\omega_1}
       \frac{w_{\omega_1}}{w_{\omega_N}}
       \prod_{k=0}^{N-1} P^{\vmu}_{\omega_k \omega_{k+1}}.
\end{align*}
If we denote by $\mathbb{P}^{\vmu}$ the path measure of the Markov chain with stationary initial distribution $\vpi^{\vmu} = \{v_i w_i\}_{i=1}^n$ and transition matrix $P^{\vmu}$, then
\[
    \mathbb{P}^{\vmu}(\omega)
    = \pi^{\vmu}_{\omega_1}
      \prod_{k=0}^{N-1} P^{\vmu}_{\omega_k \omega_{k+1}},
\]
and we may write
\begin{align*}
    e^{\Lambda^{\vc^{(2)}}(\vmu^{(2)} \mid \mP)}
    &= \lambda^{N}
       \mathbb{E}^{\vmu}\!\left[
          \frac{w_{\omega_1}\pi_{\omega_1}}{w_{\omega_N}\pi^{\vmu}_{\omega_1}}
       \right],
\end{align*}

Since we define the Markov free energy as
\[
    F^{(2)}(\vmu^{(2)} \mid \mP) = \log \lambda.
\]
We have
\[
    \Lambda^{\vc^{(2)}}(\vmu^{(2)} \mid \mP)
    = N F^{(2)}(\vmu^{(2)} \mid \mP)
      + \log \mathbb{E}^{\vmu}\!\left[
          \frac{w_{\omega_1}\pi_{\omega_1}}{w_{\omega_N}\pi^{\vmu}_{\omega_1}}
        \right].
\]

Since $w_i>0$ for all $i$, and $\pi_i>0$ for all $i$ by assumption, we introduce a Radon–Nikodym derivative
\[
    Z_N(\omega)
    = \frac{\displaystyle \frac{w_{\omega_1}\pi_{\omega_1}}{w_{\omega_N}\pi^{\vmu}_{\omega_1}}}
           {\displaystyle \mathbb{E}^{\vmu}\!\left[
                \frac{w_{\omega_1}\pi_{\omega_1}}{w_{\omega_N}\pi^{\vmu}_{\omega_1}}
             \right]},
\]
and define a new path measure $\mathbb{Q}_N$ by
\[
    \frac{\rd \mathbb{Q}_N}{\rd \mathbb{P}^{\vmu}}(\omega) = Z_N(\omega).
\]
Then
\[
    \Lambda^{\vc^{(2)}}(\vmu^{(2)} \mid \mP)
    = N F^{(2)}(\vmu^{(2)} \mid \mP)
      + \log \mathbb{E}^{\vmu}\!\left[
          \frac{w_{\omega_1}\pi_{\omega_1}}{w_{\omega_N}\pi^{\vmu}_{\omega_1}}
        \right],
\]
Differentiating the cumulant generating function, we obtain
\begin{align*}
    \nabla_{\vmu^{(2)}} \Lambda^{\vc^{(2)}}(\vmu^{(2)} \mid \mP)
    &= \frac{\mathbb{E}\!\left[\vc^{(2)} e^{\langle \vmu^{(2)}, \vc^{(2)}\rangle}\right]}
            {\mathbb{E}\!\left[e^{\langle \vmu^{(2)}, \vc^{(2)}\rangle}\right]} \\
    &= \frac{\lambda^{N}
            \mathbb{E}^{\vmu}\!\left[
               \vc^{(2)} \frac{w_{\omega_1}\pi_{\omega_1}}{w_{\omega_N}\pi^{\vmu}_{\omega_1}}
            \right]}
            {\lambda^{N}
            \mathbb{E}^{\vmu}\!\left[
               \frac{w_{\omega_1}\pi_{\omega_1}}{w_{\omega_N}\pi^{\vmu}_{\omega_1}}
            \right]} \\
    &= \mathbb{E}^{\vmu}\!\left[
               \vc^{(2)} Z_N(\omega)
            \right] = \mathbb{E}^{\mathbb{Q}_N}\!\left[\vc^{(2)}\right].
\end{align*}
In particular, since $\vc^{(2)} = N \vnu^{(2)}$, where $\vnu^{(2)}$ are empirical pair frequencies,
\[
    \frac{1}{N}\,\nabla_{\vmu^{(2)}} \Lambda^{\vc^{(2)}}(\vmu^{(2)} \mid \mP)
    = \mathbb{E}^{\mathbb{Q}_N}\!\left[\vnu^{(2)}\right].
\]

By construction, the random variable $Z_N$ depends only on the endpoints, and $\vc^{(2)}$ is the sum of $N$ transition counts. Since $\mP$ is aperiodic and irreducible, the tilted kernel $\mP^{\vmu}$ is also aperiodic and irreducible. For a long chain, the endpoints become asymptotically independent of the empirical pair counts. For a large $N$, we have the asymptotic factorization
\[
    \mathbb{P}^{\vmu}\!\bigl(\omega_1 = a,\ \omega_N = b,\ \vc^{(2)} \in \rd \vc\bigr)
    \approx (v_a w_a)(v_b w_b)\,
            \mathbb{P}^{\vmu}\!\bigl(\vc^{(2)} \in \rd \vc\bigr),
\]

Under standard ergodicity assumptions, the effect of the boundary factor $\frac{w_{\omega_1}\pi_{\omega_1}}{w_{\omega_N}\pi^{\vmu}_{\omega_1}}$ is negligible at large $N$. That is,

so $\mathbb{Q}_N$ converges to a stationary tilted Markov law. Therefore
\begin{align}
\label{eq:hessian-cov2}
    \lim_{N\to\infty} \frac{1}{N}\,
    \Lambda^{\vc^{(2)}}(\vmu^{(2)} \mid \mP)
    &= F^{(2)}(\vmu^{(2)} \mid \mP), \notag\\ 
    \lim_{N\to\infty} \frac{1}{N}\,
    \nabla_{\vmu^{(2)}} \Lambda^{\vc^{(2)}}(\vmu^{(2)} \mid \mP)
    &=  \nabla_{\vmu^{(2)}} F^{(2)}(\vmu^{(2)} \mid \mP) = \mathbb{E}^{\vmu}\!\left[\vnu^{(2)}\right], \notag\\
    \lim_{N\to\infty} \frac{1}{N}\,
    \nabla_{\vmu^{(2)}}^{\,2} \Lambda^{\vc^{(2)}}(\vmu^{(2)} \mid \mP)
    &= \nabla_{\vmu^{(2)}}^{\,2} F^{(2)}(\vmu^{(2)} \mid \mP)\notag\\
    &= \lim_{N\to\infty} \frac{1}{N}
       \left(
          \mathbb{E}^{\vmu}\left[\vc^{(2)}\left(\vc^{(2)}\right)^{T} \right]
          - \mathbb{E}^{\vmu}\!\big[\vc^{(2)}\big]\,
            \mathbb{E}^{\vmu}\!\big[\vc^{(2)}\big]^{T}
       \right)\notag\\
    &= \lim_{N\to\infty}\mathrm{Cov}^{\vmu}\!\left[\sqrt{N}\,\vnu^{(2)}\right].
\end{align}

\subsection{Information projection under Markovian measure}

Since we work with pairwise data, we introduce observations defined on transitions rather than on single states. For $k \in \mathbb{Z}_{>0}$, the observations  $\vX \colon \mathcal{S} \times \mathcal{S} \to \reals^k$, take values $\{\vx_{ij} \in \reals^{k}\}_{i,j \in \mathcal{S}}$. 

Given a path $\omega = (\omega_1,\omega_1,\dots,\omega_N)$, the sequence of observations are
\[
    \vX^{(1)},\dots,\vX^{(N-1)}, 
    \vX^{(t)} = \vx_{\omega_{t}\omega_{t+1}} \in \reals^k,
\]

We define the measurement, the random variable on the path space $\mathcal{S}^N$ as the empirical mean of observations
\[
    \vx
    = \frac{1}{N-1}\sum_{t=1}^{N-1} \vX^{(t)}.
\]

Writing this quantity in terms of the empirical pair frequencies
$\vnu^{(2)} = (\nu^{(2)}_{ij})_{i,j \in \mathcal{S}}$, we obtain the equivalent expression
\[
    \vx
    = \sum_{i,j \in \mathcal{S}} \nu^{(2)}_{ij}\,\vx_{ij}
    = \vX \vnu^{(2)},
\]

In the above short-form notation, we view $\vX$ as a linear operator acting on pairwise
distributions. The corresponding adjoint map is $\vX^{\top} \colon \reals^{k} \to \reals^{n^2}$ is defined as
\[
    (\vX^{\top} \valpha)_{ij} = \valpha^{\top} \vx_{ij},
    \qquad \valpha \in \reals^{k},
\]

\paragraph{Contracted rate function and free energy}
Using the contraction principle, the large deviation rate function for the empirical
mean $\vx$ is
\begin{align}
\label{eq:pairwise-lft-pairs}
        \phi^{(2)}(\vx \,\mid\, \mP)
    = \inf_{\vnu^{(2)} \in \riPairDelta{n}}
      \left\{
          H^{(2)}(\vnu^{(2)} \mid \mP)
          \,\big|\,
          \vX \vnu^{(2)} = \vx
      \right\}.
\end{align}

The corresponding free energy is given by the Legendre--Fenchel transform
\[
    \psi^{(2)}(\valpha\,\mid\, \mP)
    = \sup_{\vx \in \reals^{k}}
      \bigl\{
          \langle \valpha, \vx \rangle - \phi^{(2)}(\vx \,\mid\, \mP)
      \bigr\}
    = F^{(2)}(\vX^{\top} \valpha),
\]

Using the chain rule we have,
\begin{align}
\label{eq:hessian-cov3}
\nabla_{\valpha}^{2}\psi^{(2)}(\valpha\,\mid\, \mP)
&=
\vX\,\nabla_{\vmu^{(2)}}^{2}F^{(2)}(\vX^{\top}\valpha)\,\vX^{\top} \\
&=
\mathrm{Cov}^{\vX^{\top}\valpha}\!\left[\sqrt{N}\,\vx\right].
\end{align}

In a non-degenerate case where $\nabla_{\valpha}^{2}\psi^{(2)}(\valpha\,\mid\, \mP)$ is invertible, we have the relationship between $\nabla_{\valpha}^{2}\psi^{(2)}(\valpha\,\mid\, \mP)$ and $\nabla_{\vx}^{2}\phi^{(2)}(\vx \,\mid\, \mP)$ as

\[
\nabla_{\valpha}^{2}\psi^{(2)}(\valpha\,\mid\, \mP)\nabla_{\vx}^{2}\phi^{(2)}(\vx \,\mid\, \mP) = \mI.
\]

Similar to the i.i.d. case, the Hessian of the large deviation rate function $\phi^{(2)}$ is the Riemannian metric on the space of $\vx$. In equations \eqref{eq:hessian-cov1}, \eqref{eq:hessian-cov2}, and \eqref{eq:hessian-cov3}, we reinforce the choice of the metric by describing how it adapts to the variability and correlation of the data.

The optimizer of equation \eqref{eq:pairwise-lft-pairs} is of the form

\begin{align*}
    \nu^{*}_{ij} =\frac{1}{\lambda_{\mathrm{max}}\left( \mP e^{\vX^T\valpha}\right)} P_{ij}e^{\valpha^T \vx_{ij}} v_i w_j.
\end{align*}

Where  $\vv=(v_1,\cdots,v_n)$ and $\vw=(w_1,\cdots,w_n)$ are the left and right eigenvectors of matrix $\mP e^{\vX^T\valpha}$ corresponding to the principal eigenvalue $\lambda_{\mathrm{max}}\left( \mP e^{\vX^T\valpha}\right)$. And ${\vnu^*}^{(2)}$ satisfies the constraint

\begin{align*}
    \vX {\vnu^*}^{(2)} = \vx.
\end{align*}

Moreover, ${\vnu^*}^{(2)}$ is the Markov information-projection of $\mP$ on to the space space $\mathcal{A}_{\vx}^{(2)} = \{\vq^{(2)}\in \riPairDelta{n}:\vX\vq^{(2)}=\vx\}$.

\paragraph{Markov Information projection}

Information projection for
Markovian measures with pairwise empirical frequencies has been studied previously.
In particular, \cite{csiszar1987conditional} considers conditional limit theorems under
Markov conditioning and characterizes the information projection as the solution of a
variational problem: the minimizer of the Markov rate function
$H^{(2)}(\vnu^{(2)}\mid \mP)$ subject to the linear constraint
$\vX\vnu^{(2)}=\vx$.

In contrast, our approach defines information projection in the $L^{2}$ sense, following
Kolmogorov’s theory of conditional expectation. Analogous to the i.i.d.\ case, for the pairwise setting, $\Omega = \mathcal{S}^N$ and
$\mathcal{F} = 2^\Omega$. For each $\omega \in \Omega$, the empirical pair
frequency $\vnu^{(2)}(\omega)$ and empirical mean $\vx(\omega) =
\vX\vnu^{(2)}(\omega)$ are random variables on $(\Omega, \mathcal{F})$.
The empirical pair frequency takes values in the $N$-lattice pair simplex
$\riPairDelta{n}_N = \{\vq^{(2)} \in \riPairDelta{n} : N q^{(2)}_{ij} \in
\mathbb{Z}_{\geq 0}\ \forall i,j\}$, and generates the $\sigma$-algebra:
\begin{align*}
    \mathcal{F}_{\vnu^{(2)}}
    = \sigma\!\left(\left\{
        (\vnu^{(2)})^{-1}(\vq^{(2)}) : \vq^{(2)} \in \riPairDelta{n}_N
    \right\}\right)
\end{align*}
where $(\vnu^{(2)})^{-1}(\vq^{(2)}) = \{\omega \in \Omega \,:\,
\vnu^{(2)}(\omega) = \vq^{(2)}\}$. The empirical mean $\vx$ takes values in
the finite lattice of achievable means:
\begin{align*}
    \mathcal{X}^{(2)}_N
    = \{\vX\vq^{(2)} \,:\, \vq^{(2)} \in \riPairDelta{n}_N\}
    \subset \mathbb{R}^k
\end{align*}
and generates the $\sigma$-algebra:
\begin{align*}
    \mathcal{F}_{\vx}
    = \sigma\!\left(\left\{
        \vx^{-1}(\vy) \,:\, \vy \in \mathcal{X}^{(2)}_N
    \right\}\right).
\end{align*}
For $\delta > 0$, define the strictly coarser $\sigma$-algebra:
\begin{align}
\label{eq:delta-thick-sigmaalgebra-pairs}
    \mathcal{F}^\delta_{\vx}
    = \sigma\!\left(\left\{
        \vx^{-1}\!\left(B_\delta(\vy) \cap \mathcal{X}^{(2)}_N\right)
        \,:\, \vy \in \mathcal{X}^{(2)}_N
    \right\}\right)
\end{align}
where $B_\delta(\vy) = \{\vz \in \mathbb{R}^k \,:\, \|\vz - \vy\|_2 
< \delta\}$ is the open $\delta$-ball around $\vy$, giving the information
ordering:
\begin{align*}
    \mathcal{F}
    \supseteq \mathcal{F}_{\vnu^{(2)}}
    \supseteq \mathcal{F}_{\vx}
    \supseteq \mathcal{F}^\delta_{\vx}.
\end{align*}

Rather than postulating the minimization
of $H^{(2)}$ as the definition, we take the information projection of the empirical pair
frequency $\vnu^{(2)}$ onto a $\delta$-coarse sub-$\sigma$-algebra 
\[
\mathbb{E}\!\left[\vnu^{(2)} \mid \mathcal{F}^\delta_{\vx}\right].
\]
Theorem~\ref{thm:infoproj-pairs} shows that, in the $N \xrightarrow{} \infty$ limit, the $L^{2}$ projection
coincides with the variational characterization in \cite{csiszar1987conditional}. In this
sense, we recover the classical information projection for Markov chains as a consequence
of conditional expectation, rather than as a primitive definition, and thereby place
information projection for pairwise data within the standard $L^{2}$ framework of
probability theory.

\begin{theorem}[Information Projection for Pairwise Data]
\label{thm:infoproj-pairs}
Fix an arbitrarily small $\delta > 0$ and let $\mathcal{F}^\delta_{\vx}$ be
the $\delta$-coarse sub-$\sigma$-algebra as defined in
Equation~\eqref{eq:delta-thick-sigmaalgebra-pairs}.  Then as $N\to\infty$, the
information projection of $\vnu^{(2)}$ onto $\mathcal{F}^\delta_{\vx}$
satisfies:
\begin{align*}
    \mathbb{E}\!\left[\vnu^{(2)} \mid \mathcal{F}^\delta_{\vx}\right]
    = \operatorname*{arg\,inf}_{\mathcal{F}^\delta_{\vx}\text{-measurable }\vz}\,
      \mathbb{E}\!\left[\|\vz-\vnu^{(2)}\|_2^2\right]
    \xrightarrow{N\to\infty}
    \operatorname*{arg\,inf}_{\vq^{(2)} \in \tilde{\Delta}^{n}}
      \left\{
          H^{(2)}\!\left(\vq^{(2)} \mid \mP\right)
          \,\Big|\,
          \|\vX\vq^{(2)} - \vx(\omega)\|_2 < \delta
      \right\}.
\end{align*}
\end{theorem}

\begin{proof}

Fix $\vx(\omega)$ and define the $\delta$-thickened constraint set
\begin{align*}
    \mathcal{A}^{(2),\delta}_{\vx} = \left\{\vq^{(2)}\in \riPairDelta{n} :
    \|\vX\vq^{(2)} - \vx(\omega)\|_2 < \delta \right\}.
\end{align*}
For fixed $\vx(\omega) \in \mathcal{X}^{(2)}_N$, the set
$\mathcal{A}^{(2),\delta}_{\vx}$ is non-empty since there exists a sequence
$\omega$ that results in $\vx(\omega)$, and the $\delta$-thickening ensures
it has non-empty interior in $\tilde{\Delta}^{n}$, being the preimage of an
open ball under the continuous affine map $\vq^{(2)}\mapsto\vX\vq^{(2)}$.

Let $\vq_\star^{(2)}$ denote the unique minimizer of $H^{(2)}(\cdot\mid\mP)$
over $\mathcal{A}^{(2),\delta}_{\vx}$, that is
$H^{(2)}\!\left(\vq_\star^{(2)}\mid\mP\right) = \phi^{(2)}(\vx\mid\mP)$.
For $\vq^{(2)}\in\mathcal{A}^{(2),\delta}_{\vx}$, the conditional probability
satisfies
\begin{align*}
    \mathbb{P}\!\left(\vnu^{(2)}\in B_\varepsilon(\vq^{(2)})
      \,\middle|\,
      \|\vX\vnu^{(2)} - \vx(\omega)\|_2 < \delta
    \right)
    &=
    \frac{
        \mathbb{P}\!\left(\vnu^{(2)}\in B_\varepsilon(\vq^{(2)})
          \cap \mathcal{A}^{(2),\delta}_{\vx}\right)
    }{
        \mathbb{P}\!\left(\vnu^{(2)}\in\mathcal{A}^{(2),\delta}_{\vx}\right)
    }.
\end{align*}
The rate function for this conditional probability satisfies, for all
$\vq^{(2)}\in\riPairDelta{n}$ and all sufficiently small $\varepsilon > 0$:
\begin{align*}
    \lim_{N\to\infty}\frac{1}{N}\log
    \mathbb{P}\!\left(\vnu^{(2)}\in B_\varepsilon(\vq^{(2)})
      \,\middle|\,
      \|\vX\vnu^{(2)} - \vx(\omega)\|_2 < \delta
    \right)
    = -\left(
        H^{(2)}\!\left(\vq^{(2)}\mid\mP\right)
        - \phi^{(2)}\!\left(\vx\mid\mP\right)
      \right),
\end{align*}
where $H^{(2)}\!\left(\vq^{(2)}\mid\mP\right) -
\phi^{(2)}\!\left(\vx\mid\mP\right)$ is non-negative and vanishes only at
$\vq^{(2)} = \vq_\star^{(2)}$.

Let $\mathcal{B}(\riPairDelta{n})$ denote the Borel $\sigma$-algebra on
$\tilde{\Delta}^{n}$. Define the conditional pushforward measure $\mathbb{Q}_N$
on $\bigl(\tilde{\Delta}^{n},\mathcal{B}(\tilde{\Delta}^{n})\bigr)$ by
\begin{align*}
    \mathbb{Q}_N(A)
    \coloneqq
    \mathbb{P}\!\left(\vnu^{(2)}\in A
      \,\middle|\,
      \|\vX\vnu^{(2)} - \vx(\omega)\|_2 < \delta
    \right),
    \qquad A\in\mathcal{B}(\tilde{\Delta}^{n}).
\end{align*}
Let $U$ be any open set containing $\vq_\star^{(2)}$ in $\tilde{\Delta}^{n}$.
Then $U\in\mathcal{B}(\tilde{\Delta}^{n})$ and $U^c\in\mathcal{B}(\tilde{\Delta}^{n})$.
Since $H^{(2)}\!\left(\vq^{(2)}\mid\mP\right) -
\phi^{(2)}\!\left(\vx\mid\mP\right)$ is non-negative and vanishes only at
$\vq^{(2)} = \vq_\star^{(2)}$, we have
\begin{align*}
    \inf_{\vq^{(2)}\in U}
    \left(H^{(2)}\!\left(\vq^{(2)}\mid\mP\right)
    - \phi^{(2)}\!\left(\vx\mid\mP\right)\right) = 0,
    \qquad
    \inf_{\vq^{(2)}\in U^c}
    \left(H^{(2)}\!\left(\vq^{(2)}\mid\mP\right)
    - \phi^{(2)}\!\left(\vx\mid\mP\right)\right) > 0,
\end{align*}
and consequently $\mathbb{Q}_N(U^c)\to 0$, i.e.\
$\mathbb{Q}_N \Rightarrow \delta_{\vq_\star^{(2)}}$.
Since $\tilde{\Delta}^{n}$ is compact and the coordinate maps
$\vq^{(2)}\mapsto q^{(2)}_{ij}$ are bounded and continuous, weak convergence
gives
\begin{align*}
    \mathbb{E}\!\left[\vnu^{(2)} \mid \mathcal{F}^\delta_{\vx}\right]
    =
    \int_{\tilde{\Delta}^{n}} \vq^{(2)}\,\mathbb{Q}_N(\mathrm{d}\vq^{(2)})
    \;\xrightarrow{N\to\infty}\; \vq_\star^{(2)}
    = \operatorname*{arg\,inf}_{\vq^{(2)}\in\tilde{\Delta}^{n}}
      \left\{
          H^{(2)}\!\left(\vq^{(2)}\mid\mP\right)
          \,\Big|\,
          \|\vX\vq^{(2)} - \vx(\omega)\|_2 < \delta
      \right\},
\end{align*}
which is the information projection from the claim.
\end{proof}

\subsection{Geometry of the space of empirical counting frequencies $\vnu$}
\label{sec:empfreqgeom}

As described in Under a Markovian measure, the empirical counting frequencies $\vnu$ satisfy a large
deviation principle with a rate function that differs from the i.i.d.\ case. In
particular, the relevant rate function depends on the transition matrix $\mP$. We denote this rate function as
$H^{(1)}(\vnu \mid \mP)$.

Let's define $\ones_n$ as an $n$-vector of ones. When the Markov chain reduces to an i.i.d.\ process, that is, when
$\mP = \ones \vp^{T}$, for some $\vp \in \riDelta{n}$, the Markovian measure coincides with the i.i.d.\ measure $\vp$ and
the rate function reduces to the classical Sanov form:
\[
\phi^{(2)}(\vnu \mid \ones \vp^{T}) = H(\vnu \mid \vp).
\]

In the i.i.d.\ setting, the geometry on the space of empirical frequencies $\vnu$ is
independent of the underlying distribution $\vp$. This follows from the fact that,
under independence, second-order statistics are fully determined by first-order ones:
\[
\mathbb{E}[\nu_i \nu_j] = \mathbb{E}[\nu_i]\mathbb{E}[\nu_j].
\]
Hence, the metric, which depends on the second-order statistic, only depends on the empirical frequencies themselves and not on the specific value of $\vp$.

This property fails once the i.i.d.\ assumption is removed. Under a general Markovian
measure, the relation
$\mathbb{E}[\nu_i \nu_j] = \mathbb{E}[\nu_i]\mathbb{E}[\nu_j]$ no longer holds, and we may not recover second-order statistics from $\vnu$ alone. In this setting,
probability theory must be used as an extension of logic to infer the missing
second-order statistic information. 

Define the linear map $\vY \colon \reals^{n \times n} \to \reals^n$ by 
\[
(\vY\,\vnu^{(2)})_i \;=\; \sum_{j=1}^n \nu^{(2)}_{ij}, \qquad i=1,\dots,n.
\]
The pairwise counting frequency $\vnu^{(2)}$ and the empirical frequency $\vnu$ are related via the map $\vnu = \vY \vnu^{(2)}$.

To characterize the geometry at a given empirical frequency $\vq$, let $\vq$ and
$\valpha$ form a Legendre-Fenchel pair. The pairs satisfy the relation $\vq = \vpi^{{\vY}^T\alpha}$, where $\vpi^{{\vY}^T\alpha}$ is the stationary distribution of the Markov chain with transition matrix $\mP^{{\vY}^T\alpha}$.

\begin{lemma}
    \label{thm:markov-metric}
     Let $\vq \in \riDelta{n}$ and $\valpha \in \reals^n$ be a Legendre-Fenchel pair, then under the stationary ergodic Markovian measure $\mP^{\vY^T \valpha}$, the limit of the following Covariance matrix,

     \begin{align*}
        \lim_{N\to\infty} N\,\mathrm{Cov}^{\vY^T \valpha}(\vnu)
        =
        D_{\vq}(\mI-\mP^{\vY^T \valpha}+\ones\vq^T)^{-1}
        +
        \Big(D_{\vq}(\mI-\mP^{\vY^T \valpha}+\ones\vq^T)^{-1}\Big)^{\top}
        -
        D_{\vq}( \mI+\ones\vq^T )
    \end{align*}
where $D_{\vq}$ is the diagonal matrix with $\vq$ as entries.
\end{lemma}

\begin{proof}

The proof is in section \ref{appendix:proof} of the Appendix.

\end{proof}

The simplex $\riDelta{n}$ is an $ n-1$-dimensional manifold, and one only needs $n-1$ entries of $\vq$ to chart the manifold. For example, if $\vnu_{-1} =[nu_1,\nu_2,\ldots,\nu_{n-1}]^\top$ denotes $\vnu$ excluding the last entry, and similarly $\vq_{-1} = [q_1,q_2,\ldots,q_{n-1}]^T$, then the metric at $\vq_{-1}$ is 
\[
\left( \lim_{N\to\infty} N\,\mathrm{Cov}^{\vY^T \valpha}(\vnu_{-1}) \right)^{-1}.
\]

And according to Lemma \ref{thm:markov-metric}, the Covariance depends on the specific value of the underlying Markovian measure $\mP$.

The following example demonstrates the metric's dependence on the underlying values of the Markov transition probability matrix.

\begin{example}
Let $n\ge 2$ and fix a point $\vq=\vpi\in \riDelta{n}$. 
For $\epsilon\in(0,1]$ consider the transition matrix
\[
\mP(\epsilon)\coloneqq (1-\epsilon)\mI+\epsilon\,\Pi .
\]
Then $\vpi$ is stationary for every $\epsilon$, since $\vpi^{\top}\mP(\epsilon)=\vpi^{\top}$.

At the Legendre--Fenchel dual point $\valpha=0$, we have $\mP^{\vY^{\top}\valpha}=\mP(\epsilon)$ and Lemma~\ref{thm:markov-metric} gives
\begin{align*}
\lim_{N\to\infty}N\,\mathrm{Cov}(\vnu)
&=
D_{\vpi}\bigl(\mI-\mP(\epsilon)+\Pi\bigr)^{-1}
+
\Big(D_{\vpi}\bigl(\mI-\mP(\epsilon)+\Pi\bigr)^{-1}\Big)^{\top}
-
D_{\vpi}(\mI+\Pi).
\end{align*}
Compute
\[
\mI-\mP(\epsilon)+\Pi
=
\epsilon \mI+(1-\epsilon)\Pi,
\qquad
\bigl(\epsilon \mI+(1-\epsilon)\Pi\bigr)^{-1}
=
\Pi+\frac{1}{\epsilon}(\mI-\Pi),
\]
so
\begin{align*}
\lim_{N\to\infty}N\,\mathrm{Cov}(\vnu)
&=
\left(\frac{2}{\epsilon}-1\right)\Bigl(D_{\vpi}-\vpi\vpi^{\top}\Bigr)
=
\frac{2-\epsilon}{\epsilon}\Bigl(D_{\vpi}-\vpi\vpi^{\top}\Bigr).
\end{align*}

We use the coordinates $\vnu_{-1}=(\nu_1,\dots,\nu_{n-1})^{\top}$ as a chart for $\riDelta{n}$ and are interested in the metric at $\vpi_{-1}=(\pi_1,\dots,\pi_{n-1})^{\top}$. Then
\[
\lim_{N\to\infty}N\,\mathrm{Cov}(\vnu_{-1})
=
\frac{2-\epsilon}{\epsilon}\Bigl(D_{\vpi_{-1}}-\vpi_{-1}\vpi_{-1}^{\top}\Bigr)
\]
Since $\pi_n=1-\ones^{\top}\vpi_{-1}>0$, the matrix $D_{\vpi_{-1}}-\vpi_{-1}\vpi_{-1}^{\top}$ is invertible and
\[
\bigl(D_{\vpi_{-1}}-\vpi_{-1}\vpi_{-1}^{\top}\bigr)^{-1}
=
D_{\vpi_{-1}}^{-1}
+
\frac{1}{\pi_n}\,\ones\ones^{\top}.
\]
Therefore, the metric as a function of $\epsilon$ at $\vpi \in \riPairDelta{n}$ using $(n-1)$ coordinate chart is
\begin{align*}
\mG(\epsilon)
&\coloneqq
\Bigl(\lim_{N\to\infty}N\,\mathrm{Cov}(\vnu_{-n})\Bigr)^{-1}
=
\frac{\epsilon}{2-\epsilon}
\left(
D_{\vpi_{-1}}^{-1}
+
\frac{1}{\pi_n}\,\ones\ones^{\top}
\right).
\end{align*}
\end{example}

\subsection{Curvature of empirical pair-counting frequencies}
\label{sec:curvature}
It is well known that the simplex $\riDelta{n}$ equipped with the Fisher-Rao metric has a spherical geometry \cite{Rao1992-qj}. In other words, the space of empirical frequency under the assumption of i.i.d. has a spherical geometry and has a constant positive Riemannian curvature tensor. This is because the Hessian of $H(\vnu \,\mid\, \vp)$ is the Fisher-Rao metric. We investigate the curvature of the empirical pairwise counting frequency $\riPairDelta{n}$ under the Markovian assumption. Although the empirical pair-counting frequencies occupy a space similar to that of the probability simplex, the geometry is not spherical. We consider the case of $n=2$  and state the following theorem.

\begin{theorem}
\label{thm:curvature}
    Under the assumption of a stationary Markovian measure with transition probability matrix $\mP$, the Gaussian curvature at a point $\vq = (q_{11},q_{12},q_{21},q_{22}) \in \riPairDelta{2}$ is 

    \begin{align}
        K(\vq) = \frac{q_{1}q_{2}}{q_{12}}.
    \end{align}
    Where 
    \[
    q_1 = q_{11} + q_{12}, \qquad q_2 = q_{22} + q_{21}.
    \]

    \begin{proof}
    
        The proof is in section \ref{proof:curvature} of the Appendix.
    \end{proof}
\end{theorem}

%%%%%%%%%%%%%%%%%% CONCLUSION %%%%%%%%%%%%%%%%%%
\section{Conclusion}
\label{sec:4}
The present work pays particular attention to the critical distinction between empirical counting frequencies under i.i.d. and probability distributions. It utilizes concrete fundamentals of information theory to establish a framework for a broader information geometry that encompasses both data and models on an equal footing.  According to Kolmogorov's theory of probability, empirical frequencies are themselves random variables with an associated probability. The distinction is essential in applying large deviation theory results as a limit theorem.  Using large deviation theory, we revealed a rich structure on the manifold of empirical data under i.i.d. samples. We also distinguish the manifold of empirical data from the manifold of statistical models in information geometry literature: the manifold of probability distributions. Heuristically, the manifold of empirical data {{ {\em ad infinitum} is a ``dual'' representation }} to the space of statistical models. The Legendre-Fenchel transform, central to Gibbs' statistical thermodynamics method, large deviation theory, and convex optimization, provided us with a dual space of internal energies. The energetic description of data provides a certain additivity in energy space, and the description of mean internal energy, free energy, and entropy allows us to formulate the very {\it statistical physics}. The dual notion of empirical frequencies and internal energies bridge the methodologies used by information theorists and physicists. We provide a robust idea of information projection using probability theory.

The presentation of our methods is limited to a discrete $\Omega$ space, and with the language of Polish spaces from large deviation theory and real analysis, our insights can easily extend to continuous spaces like $\reals^n$. Instead of focusing on the differential geometry topics of tangent spaces, affine connections, geodesics, and the Riemannian curvature, we focus on justifying the use of the Hessian of the rate function (also called the Fisher information matrix) as an appropriate metric in Sec. \ref{subsection:contraction}. The notion of dually flat connections in information geometry naturally arises because our data manifold possesses a global chart in both empirical frequency and energy coordinates. Our treatment pivots from the space of a statistical model in IG to a space of statistical measurements.  In engineering and applied mathematics, which deal with data and quantitative measurements, numerals are empirically given {\em a priori}, suggesting a natural chart, at least locally, for any geometric modeling of statistical data. The analytical definition of convexity then follows. Furthermore, a proper Riemannian metric should not be based on the local Euclidean space and Lebesgue measure of the numerals but account for statistical uncertainties within the data encoded in entropy functions.

%%%%%%%%%%%%%%%%%% BIBLIOGRAPHY %%%%%%%%%%%%%%%%%%

\input{main.bbl}

% \bibliography{reference}
%%%%%%%%%%%%%%%%%% APPENDIX %%%%%%%%%%%%%%%%%%

\appendix

\subsection{Proof of Lemma \ref{thm:markov-metric}}
\label{appendix:proof}
\begin{proof}

Throughout this proof, we work under the stationary tilted Markov law
with transition matrix $\mP^{\vY^{\top}\valpha}$ and initial distribution
$\vpi^{\vY^{\top}\valpha}$. We will omit the superscript ${\vY^{\top}\valpha}$ everywhere in this proof for simplicity.
    The associated covariance is
$\lim_{N \to \infty}\mathrm{Cov}\!\left[\sqrt{N}\,\vnu\right] = \lim_{N \to \infty} \mathrm{Cov}\!\left[\vnu\right]$.

Componentwise,
\begin{align*}
N\,\mathrm{Cov}(\nu_i,\nu_j)
=
N\Big(
\mathbb{E}[\nu_i\nu_j]
-
\mathbb{E}[\nu_i]
\mathbb{E}[\nu_j]
\Big).
\end{align*}

Write the empirical frequencies in terms of the path $\omega=(\omega_1,\dots,\omega_N)$ using the indicator $\indicator_i(\omega_j)\coloneqq \delta_i^{\omega_j}$:
\begin{align*}
\nu_i(\omega)
=
\frac{1}{N}
\sum_{k=1}^N
\indicator_i(\omega_k),
\qquad
\nu_j(\omega)
=
\frac{1}{N}
\sum_{\ell=1}^N
\indicator_j(\omega_\ell).
\end{align*}

Then,

\begin{align*}
N\Big(
\mathbb{E}[\nu_i\nu_j]
-
\mathbb{E}[\nu_i]\,
\mathbb{E}[\nu_j]
\Big)
&=
\frac{1}{N}
\sum_{k=1}^N\sum_{\ell=1}^N
\mathrm{Cov}
\!\left(
\indicator_i(\omega_k),
\indicator_j(\omega_\ell)
\right) \\
&=
\frac{1}{N}
\sum_{k=1}^N
\mathrm{Cov}
\!\left(
\indicator_i(\omega_k),
\indicator_j(\omega_k)
\right) \\
&\quad
+
\frac{1}{N}
\sum_{k=1}^{N}
\sum_{b=1}^{N-k}
\mathrm{Cov}
\!\left(
\indicator_i(\omega_k),
\indicator_j(\omega_{k+b})
\right)\\
&\quad
+
\frac{1}{N}
\sum_{k=1}^{N}
\sum_{b=1}^{N-k}
\mathrm{Cov}
\!\left(
\indicator_j(\omega_k),
\indicator_i(\omega_{k+b})
\right) 
\end{align*}

Taking the limit as $N\to\infty$ and using the stationary property of Markov chain, we have
\begin{align*}
\lim_{N\to\infty}
N\,\mathrm{Cov}(\nu_i,\nu_j)
&=
\sum_{b=1}^{\infty}
\mathrm{Cov}
\!\left(
\indicator_i(\omega_1),
\indicator_j(\omega_{1+b})
\right) \\
&\quad
+
\sum_{b=1}^{\infty}
\mathrm{Cov}
\!\left(
\indicator_j(\omega_1),
\indicator_i(\omega_{1+b})
\right)
+
\mathrm{Cov}
\!\left(
\indicator_i(\omega_1),
\indicator_j(\omega_1)
\right).
\end{align*}

Define the diagonal matrix of stationary probabilities by
\[
D_{\vpi} \coloneqq \mathrm{diag}(\vpi),
\]
and define the rank-one matrix with each row equal to $\vpi^{\top}$ by
\[
\Pi \coloneqq \ones \vpi^{\top},
\]

For $b\ge 0$,
\[
\mathbb{P}(\omega_1=i,\omega_{1+b}=j)=\pi_i(\mP^{b})_{ij},
\qquad
\mathbb{P}(\omega_1=i)=\pi_i,
\qquad
\mathbb{P}(\omega_{1+b}=j)=\pi_j,
\]
hence
\begin{align*}
\mathrm{Cov}\!\left(\indicator_i(\omega_1),\indicator_j(\omega_{1+b})\right)
&=
\mathbb{P}(\omega_1=i,\omega_{1+b}=j)-\mathbb{P}(\omega_1=i)\mathbb{P}(\omega_{1+b}=j) \\
&=
\pi_i(\mP^{b})_{ij}-\pi_i\pi_j.
\end{align*}

We rewrite
\[
\pi_i(\mP^{b})_{ij}-\pi_i\pi_j
=
\pi_i\bigl(\mP^{b}-\Pi\bigr)_{ij},
\]
since $\Pi=\ones\vpi^{\top}$ satisfies $\Pi_{ij}=\pi_j$.

Moreover, using $\mP\Pi=\Pi$ and $\Pi\mP=\Pi$, we obtain
\[
\mP^{b}-\Pi
=
(\mP-\Pi)^{b},
\qquad b\ge 1.
\]

For $b\ge 1$ we have
\begin{align*}
\mathrm{Cov}\!\big(\indicator_i(\omega_1),\indicator_j(\omega_{1+b})\big)
&=
\mathbb{P}(\omega_1=i,\omega_{1+b}=j)-\mathbb{P}(\omega_1=i)\mathbb{P}(\omega_{1+b}=j)\\
&=
\pi_i(\mP^{b})_{ij}-\pi_i\pi_j
=
\bigl(D_{\vpi}(\mP^{b}-\Pi)\bigr)_{ij}
=
\bigl(D_{\vpi}(\mP-\Pi)^{b}\bigr)_{ij},
\end{align*}

For the swapped term,
\begin{align*}
\mathrm{Cov}\!\big(\indicator_j(\omega_1),\indicator_i(\omega_{1+b})\big)
&=
\bigl(D_{\vpi}(\mP-\Pi)^{b}\bigr)_{ji}
=
\bigl(\bigl(D_{\vpi}(\mP-\Pi)^{b}\bigr)^{\top}\bigr)_{ij}.
\end{align*}

Therefore the full asymptotic covariance matrix can be written as
\begin{align*}
\lim_{N\to\infty} N\,\mathrm{Cov}(\vnu)
&=
\sum_{b=1}^{\infty} D_{\vpi}(\mP-\Pi)^{b}
\;+\;
\sum_{b=1}^{\infty} \bigl(D_{\vpi}(\mP-\Pi)^{b}\bigr)^{\top}
\;+\;
D_{\vpi}(\mI - \Pi).
\end{align*}

We note that the series above converges because $\mP-\Pi$ has spectral radius strictly less than one under the  ergodicity assumption,
\begin{align*}
\rho(\mP-\Pi)<1,
\end{align*}
so $(\mP-\Pi)^{b}\to 0$ as $b\to\infty$ and the matrix geometric series
$\sum_{b=1}^{\infty}(\mP-\Pi)^{b}$ converges absolutely.

The following series converges to
\begin{align*}
\sum_{b=1}^{\infty}(\mP-\Pi)^{b}
&=
\sum_{b=0}^{\infty}(\mP-\Pi)^{b} - \mI
=
(\mI-(\mP-\Pi))^{-1} -\mI
\end{align*}

Therefore,
\begin{align*}
\lim_{N\to\infty} N\,\mathrm{Cov}(\vnu)
&=
D_{\vpi}(\mI - \Pi)
+
D_{\vpi}\Big((\mI-\mP+\Pi)^{-1}-\mI\Big)
+
\Big(D_{\vpi}\big((\mI-\mP+\Pi)^{-1}-\mI\big)\Big)^{\top} \\
&=
D_{\vpi}(\mI-\mP+\Pi)^{-1}
+
\Big(D_{\vpi}(\mI-\mP+\Pi)^{-1}\Big)^{\top}
-
D_{\vpi}( \mI+\Pi ).
\end{align*}

Replacing $\mP$ with $\mP^{\vY^T \valpha}$ and $\vpi$ with $\vq$ finishes the proof.

\end{proof}

\subsection{Proof of theorem \ref{thm:curvature}}
\label{proof:curvature}
For $n=2$, the space of pairwise frequencies $\riPairDelta{2}$ is $2$-dimensional. 
Fix $(\vnu_{11}, \vnu_{22})= (a,b)$ as independent variables then
\[
\vnu_{12}= \vnu_{21} = \frac{1-a-b}{2}.
\]
The large-deviation rate function is

\begin{align*}
H(a,b)
&= \sum_{i,j=1}^2 \vnu_{ij}\log \left( \frac{\vnu_{ij}}{P_{ij}\sum_k \vnu_{ik}} \right)\\
&= a \log \left(\frac{a}{P_{11}\frac{1+a-b}{2}}\right)
+ \frac{1-a-b}{2}\log \left(\frac{\frac{1-a-b}{2}}{P_{12}\frac{1+a-b}{2}}\right)\\
&\quad + \frac{1-a-b}{2}\log \left(\frac{\frac{1-a-b}{2}}{P_{21}\frac{1-a+b}{2}}\right)
+ b \log \left(\frac{b}{P_{22}\frac{1-a+b}{2}}\right).
\end{align*}

\begin{align*}
H(a,b)
&= a\log a + (1-a-b)\log(1-a-b) + b\log b\\
&\quad - \frac{1+a-b}{2}\log(1+a-b)
- \frac{1-a+b}{2}\log(1-a+b)\\
&\quad - a\log P_{11}
- \frac{1-a-b}{2}\log(P_{12}P_{21})
- b\log P_{22}.
\end{align*}

Computing the hessian gives us the entries of the metric of as follows

\[
H_{aa}
=
\dfrac{1}{a}
+\dfrac{1}{1-a-b}
-\dfrac{1}{2(1+a-b)}
-\dfrac{1}{2(1-a+b)},
\]

\[
H_{ab}=H_{ba}
=
\dfrac{1}{1-a-b}
+\dfrac{1}{2(1+a-b)}
+\dfrac{1}{2(1-a+b)},
\]

\[
H_{bb}
=
\dfrac{1}{b}
+\dfrac{1}{1-a-b}
-\dfrac{1}{2(1+a-b)}
-\dfrac{1}{2(1-a+b)}.
\]

To simplify the notation, define
\[
s=1-a-b,\qquad u=1+a-b,\qquad v=1-a+b.
\]
The entries of the metric are
\[
g_{aa}=H_{aa}
=
\frac{1}{a}+\frac{1}{s}-\frac{1}{2u}-\frac{1}{2v},
\]
\[
g_{ab}=g_{ba}=H_{ab}
=
\frac{1}{s}+\frac{1}{2u}+\frac{1}{2v},
\]
\[
g_{bb}=H_{bb}
=
\frac{1}{b}+\frac{1}{s}-\frac{1}{2u}-\frac{1}{2v}.
\]

The determinant of the Hessian metric is
\[
\det(H)=H_{aa}H_{bb}-H_{ab}^2.
\]

Set
\[
X=\frac1s-\frac1{2u}-\frac1{2v},
\qquad
Y=\frac1s+\frac1{2u}+\frac1{2v}.
\]
Then
\[
H_{aa}=\frac1a+X,\qquad H_{bb}=\frac1b+X,\qquad H_{ab}=Y.
\]
Hence
\begin{align*}
\det(H)
&=\left(\frac1a+X\right)\left(\frac1b+X\right)-Y^2\\
&=\frac1{ab}+\frac{X}{a}+\frac{X}{b}+X^2-Y^2.
\end{align*}

Now,
\[
X^2-Y^2=(X-Y)(X+Y).
\]
Since
\[
X-Y
=
\left(\frac1s-\frac1{2u}-\frac1{2v}\right)
-
\left(\frac1s+\frac1{2u}+\frac1{2v}\right)
=
-\frac1u-\frac1v,
\]
and
\[
X+Y
=
\left(\frac1s-\frac1{2u}-\frac1{2v}\right)
+
\left(\frac1s+\frac1{2u}+\frac1{2v}\right)
=
\frac{2}{s},
\]
we obtain
\[
X^2-Y^2
=
\left(-\frac1u-\frac1v\right)\frac{2}{s}
=
-\frac{2}{su}-\frac{2}{sv}.
\]

Also,
\[
\frac{X}{a}+\frac{X}{b}
=
\left(\frac1a+\frac1b\right)
\left(\frac1s-\frac1{2u}-\frac1{2v}\right).
\]
Therefore
\begin{align*}
\det(H)
&=
\frac1{ab}
+
\left(\frac1a+\frac1b\right)
\left(\frac1s-\frac1{2u}-\frac1{2v}\right)
-\frac{2}{su}-\frac{2}{sv}\\
&=
\frac1{ab}
+\frac1{as}+\frac1{bs}
-\frac1{2au}-\frac1{2av}
-\frac1{2bu}-\frac1{2bv}
-\frac{2}{su}-\frac{2}{sv}.
\end{align*}

Group the terms with denominator $u$:
\[
-\frac1{2au}-\frac1{2bu}-\frac{2}{su}
=
-\frac1u\left(\frac1{2a}+\frac1{2b}+\frac{2}{s}\right),
\]
and similarly the terms with denominator $v$:
\[
-\frac1{2av}-\frac1{2bv}-\frac{2}{sv}
=
-\frac1v\left(\frac1{2a}+\frac1{2b}+\frac{2}{s}\right).
\]
Thus
\[
\det(H)
=
\frac1{ab}+\frac1{as}+\frac1{bs}
-\frac1u\left(\frac1{2a}+\frac1{2b}+\frac{2}{s}\right)
-\frac1v\left(\frac1{2a}+\frac1{2b}+\frac{2}{s}\right).
\]

Using
\[
\frac1u+\frac1v=\frac{u+v}{uv}=\frac{2}{uv},
\]
we get
\[
\det(H)
=
\frac1{ab}+\frac1{as}+\frac1{bs}
-\frac{2}{uv}\left(\frac1{2a}+\frac1{2b}+\frac{2}{s}\right).
\]
Hence
\[
\det(H)
=
\frac1{ab}+\frac1{as}+\frac1{bs}
-\frac1{auv}-\frac1{buv}-\frac4{suv}.
\]

Putting everything over the common denominator $absuv$ gives
\[
\det(H)
=
\frac{suv+buv+auv-bs-av-4ab}{absuv}.
\]
Since
\[
s+a+b=1,
\]
the first three terms combine as
\[
suv+buv+auv=uv(s+a+b)=uv,
\]
so the numerator becomes
\[
uv-s(a+b)-4ab.
\]
Now
\[
uv=(1+a-b)(1-a+b)=1-(a-b)^2=1-a^2+2ab-b^2,
\]
and
\[
s(a+b)=(1-a-b)(a+b)=a+b-a^2-2ab-b^2.
\]
Therefore
\begin{align*}
uv-s(a+b)-4ab
&=(1-a^2+2ab-b^2)-(a+b-a^2-2ab-b^2)-4ab\\
&=1-a-b\\
&=s.
\end{align*}
Thus
\[
\det(H)=\frac{s}{absuv}=\frac{1}{abuv}.
\]

Since we utilize a Hessian metric, the Gaussian curvature is the following scalar \cite{duistermaat1999hessian},
\[
K
=
\frac{
- H_{bb}\bigl(H_{aaa}H_{abb}-H_{aab}^{2}\bigr)
+ H_{ab}\bigl(H_{aaa}H_{bbb}-H_{aab}H_{abb}\bigr)
- H_{aa}\bigl(H_{aab}H_{bbb}-H_{abb}^{2}\bigr)
}{
4\bigl(H_{aa}H_{bb}-H_{ab}^{2}\bigr)^{2}
}.
\]

Here,
\[
H_{aa}:=\partial_{a}^{2}H,\qquad
H_{ab}:=\partial_{a}\partial_{b}H,\qquad
H_{bb}:=\partial_{b}^{2}H,
\]
and
\[
H_{aaa}:=\partial_{a}^{3}H,\qquad
H_{aab}:=\partial_{a}^{2}\partial_{b}H,\qquad
H_{abb}:=\partial_{a}\partial_{b}^{2}H,\qquad
H_{bbb}:=\partial_{b}^{3}H.
\]

Equivalently, in a determinant form as
\[
K
=
-\frac{1}{4\bigl(H_{aa}H_{bb}-H_{ab}^{2}\bigr)^{2}}
\det
\begin{pmatrix}
H_{aa} & H_{ab} & H_{bb}\\
H_{aaa} & H_{aab} & H_{abb}\\
H_{aab} & H_{abb} & H_{bbb}
\end{pmatrix}.
\]

The next step is to substitute the second and third-order derivatives of \(H\) into
\[
\det
\begin{pmatrix}
H_{aa} & H_{ab} & H_{bb}\\
H_{aaa} & H_{aab} & H_{abb}\\
H_{aab} & H_{abb} & H_{bbb}
\end{pmatrix}.
\]

Recall that
\[
H_{aa}
=
\frac{1}{a}
+\frac{1}{s}
-\frac{1}{2u}
-\frac{1}{2v},
\]
\[
H_{ab}=H_{ba}
=
\frac{1}{s}
+\frac{1}{2u}
+\frac{1}{2v},
\]
\[
H_{bb}
=
\frac{1}{b}
+\frac{1}{s}
-\frac{1}{2u}
-\frac{1}{2v}.
\]

Their third derivatives are

\[
H_{aaa}
=
-\frac{1}{a^{2}}
+\frac{1}{s^{2}}
+\frac{1}{2u^{2}}
-\frac{1}{2v^{2}},
\]
\[
H_{aab}
=
\frac{1}{s^{2}}
-\frac{1}{2u^{2}}
+\frac{1}{2v^{2}},
\]
\[
H_{abb}
=
\frac{1}{s^{2}}
+\frac{1}{2u^{2}}
-\frac{1}{2v^{2}},
\]
\[
H_{bbb}
=
-\frac{1}{b^{2}}
+\frac{1}{s^{2}}
-\frac{1}{2u^{2}}
+\frac{1}{2v^{2}}.
\]

We found the determinant using SymPy's symbolic computation \cite{meurer2017sympy}. The script is available at \cite{muppirala_2026_19361932}.

\[
\det
\begin{pmatrix}
H_{aa} & H_{ab} & H_{bb}\\
H_{aaa} & H_{aab} & H_{abb}\\
H_{aab} & H_{abb} & H_{bbb}
\end{pmatrix} = \frac{-2}{uvs a^2 b^2}.
\]

Hence, the Gaussian curvature is
\begin{align}
    K = -\frac1{4} (abuv)^2\left(  \frac{-2}{uvs a^2 b^2}\right) = \frac{1}{2} \frac{uv}{s}.
\end{align}

In terms of counting frequencies, we have  
\[\nu_1 = \nu_{11} + \nu_{12} = a + \frac{1-a-b}{2} = \frac{u}{2} \qquad \nu_2 = \nu_{22} + \nu_{21} = b + \frac{1-a-b}{2} = \frac{v}{2} \qquad \nu_{12} =\nu_{21} = \frac{1-a-b}{2} = \frac{s}{2}.\]

So, the Gaussian curvature at a point $\vq \in \riPairDelta{2}$ is
\begin{align}
    K(\vq) = \frac{q_1q_2}{q_{12}}.
\end{align}

\end{document}

%% file: macros.tex
\def\rd{{\rm d}}

\def\vc{{\bf c}}

\def\vp{{\bf p}}
\def\vq{{\bf q}}

\def\vx{{\bf x}}
\def\vy{{\bf y}}

\def\vv{{\bf v}}
\def\vw{{\bf w}}
\def\vz{{\bf z}}
\def\vnu{\boldsymbol{\nu}}
\def\vmu{\boldsymbol{\mu}}

\def\vpi{\boldsymbol{\pi}}

\def\mG{{\bf G}}

\def\mI{{\bf I}}

\def\mP{{\bf P}}

\def\vX{{\bf X}}
\def\vY{{\bf Y}}

\newcommand{\Prob}{\mathbb{P}}
\newcommand{\bmat}[1]{\begin{bmatrix}#1\end{bmatrix}}

\newcommand{\reals}{\mathbb{R}}
\def\valpha{{\boldsymbol{\alpha}}}
\def\indicator{{\mathbbm{1}}}

\newcommand{\Conv}[1]{\boldsymbol{\mathrm{Conv}}\left( #1 \right)}
\newcommand{\ones}{\mathbf{1}}
\newcommand{\riDelta}[1]{\operatorname{ri}(\Delta^{#1})}
\newcommand{\riPairDelta}[1]{\operatorname{ri}(\widetilde{\Delta}^{#1})}

\newtheorem{theorem}{Theorem}
\renewenvironment{proof}[1][\proofname]{{\bfseries #1.}}{\qed}
\newtheorem{corollary}[theorem]{Corollary}
\newtheorem{lemma}[theorem]{Lemma}
\newtheorem{definition}{Definition}
\newtheorem{notation}{Notation}
\newtheorem{example}{Example}
\newtheorem*{remark}{Remark}